\documentclass[lettersize,journal]{IEEEtran}
\usepackage{amsmath,amsfonts}
\usepackage{algorithmic}
\usepackage{algorithm}
\newlength \figwidth
\usepackage{array}
\usepackage{booktabs}
\usepackage[caption=false,font=footnotesize,labelfont=rm,textfont=rm]{subfig}
\usepackage{textcomp}
\usepackage{stfloats}
\usepackage{url}
\usepackage{verbatim}
\usepackage{graphicx}
\usepackage{float} 
\usepackage{cite}
\usepackage{amsthm}
\usepackage{amssymb}
\usepackage{xcolor}
\usepackage{dsfont}
\usepackage{soul}
\usepackage{makecell}
\usepackage{bbding}
\def\snr{\mathtt{SNR}}

\definecolor{bittersweet}{rgb}{1.0, 0.44, 0.37}
\definecolor{glaucous}{rgb}{0.38, 0.51, 0.71}
\definecolor{gainsboro}{rgb}{0.86, 0.86, 0.86}
\definecolor{babyblueeyes}{rgb}{0.63, 0.79, 0.95}
\definecolor{silver}{rgb}{0.75, 0.75, 0.75}
\definecolor{neoncarrot}{rgb}{1.0, 0.64, 0.26}
\definecolor{Gray}{gray}{0.6}
\definecolor{LightCyan}{rgb}{0.88,1,1}
\definecolor{BackgroundLightBlue}{rgb}{0.97,0.97,1}
\definecolor{BackgroundGray}{gray}{0.98}

\newtheorem{Problem}{\bf Problem}
\newtheorem{Proposition}{\bf Proposition}

\newtheorem{Remark}{\bf Remark}

\newcommand{\blue}[1]{\textcolor{black}{#1}}

\def\nb0{{\mathbf{0}}}
\def\nb1{{\mathbf{1}}}

\def\ncalB{{\mathcal{B}}}

\def\ncalF{{\mathcal{F}}}
\def\ncalG{{\mathcal{G}}}

\def\ncalP{{\mathcal{P}}}

\def\snr{\mathtt{SNR}}

\begin{document}

\title{Wireless Edge Content Broadcast via Integrated Terrestrial and Non-terrestrial Networks}

\author{Feng Wang,~\IEEEmembership{Member,~IEEE,} Giovanni Geraci,~\IEEEmembership{Senior Member,~IEEE,}\\Lingxiang Li, Peng Wang, and Tony Q. S. Quek,~\IEEEmembership{Fellow,~IEEE}
\thanks{Feng Wang, Peng Wang, and Tony Q. S. Quek are with the Information System Technology and Design Pillar, Singapore University of Technology and Design, Singapore 487372 (e-mail: feng2\_wang@sutd.edu.sg; peng\_wang2@sutd.edu.sg; tonyquek@sutd.edu.sg).}
\thanks{Giovanni~Geraci is with Telef\'{o}nica Research and Universitat Pompeu Fabra (UPF), Barcelona, Spain (e-mail: giovanni.geraci$@$upf.edu).}
\thanks{Lingxiang Li is with National Key Laboratory of Wireless Communications, University of Electronic Science and Technology of China, Chengdu 611731, China (e-mail: lingxiang.li$@$uestc.edu.cn).}
\thanks{F.~Wang and T.~Q.~S.~Quek were supported in part by the National Research Foundation, Singapore and Infocomm Media Development Authority under its Future Communications R\&D Programme.} 
\thanks{G.~Geraci was supported in part by the Spanish Research Agency through grants PID2021-123999OB-I00, CEX2021-001195-M, and CNS2023-145384, by the UPF-Fractus Chair, by the Spanish Ministry of Economic Affairs and Digital Transformation and the European Union NextGenerationEU project UNICO 5G I+D, and by the HORIZON-SESAR-2023-DES-ER-02 project ANTENNAE (101167288).}
\thanks{Some of these results were presented at IEEE Globecom 2023 \cite{WanGerQue2023}.}
}


\maketitle

\pagestyle{empty}  
\thispagestyle{empty} 

\begin{abstract}

Non-terrestrial networks (NTN) have emerged as a transformative solution to bridge the digital divide and deliver essential services to remote and underserved areas. In this context, low Earth orbit (LEO) satellite constellations offer remarkable potential for efficient cache content broadcast in remote regions, thereby extending the reach of digital services. 
In this paper, we introduce a novel approach to optimize wireless edge content placement using NTN. 
\blue{Despite wide coverage, the varying NTN transmission capabilities must be carefully aligned with each content placement to maximize broadcast efficiency.
In this paper, we introduce a novel approach to optimize wireless edge content placement using NTN, positioning NTN as a complement to TN for achieving optimal content broadcasting.}
Specifically, we dynamically select content for placement via NTN links. This selection is based on popularity and suitability for delivery through NTN, while considering the orbital motion of LEO satellites. Our system-level case studies, based on a practical LEO constellation, demonstrate the significant improvement in placement speed compared to existing methods, which neglect network mobility. 
We also demonstrate that NTN links significantly outperform standalone wireless TN solutions, particularly in the early stages of content delivery. This advantage is amplified when there is a higher correlation of content popularity across geographical regions.
\end{abstract}


\section{Introduction}

\subsection{Background and Motivation}

\blue{In our increasingly interconnected world, where seamless access to information has become a necessity, the demand for network connectivity that enables new services is rapidly increasing}~\cite{OUGHTON2024102766,WanYouGao2023,garcia2021ieee,GioGerCar2023,AnSunLin2024}. 
While optical fiber links have long been the backbone of high-speed data transmission, their deployment in remote and underserved areas can be logistically challenging and financially impractical. Enter non-terrestrial networks (NTN), a transformative solution that promises to bridge the digital divide and enable disruptive use cases by leveraging airborne or space-based infrastructure \cite{JiaWanLv2023,giordani2020non,GerLopBen2022,kodheli2020satellite,AzaSolCha2022,KarKhoAlf2021,GuidottiVCACMEA19}. With the ability to provide connectivity to even the most remote corners of our planet, NTN represents a paradigm shift in delivering essential services to areas that were previously beyond the reach of traditional terrestrial networks \cite{RinMaaTor2020,DarKurYan2022,BenGerLop2022,GerGarAza2022,LeySorMat2022,LuYanPap2023}.

Within the realm of NTN, an emerging application with remarkable potential is the utilization of low Earth orbit (LEO) satellite constellations for efficient cache content broadcast to remote areas \cite{ZhaZhaAn2022,RinMaaTor2021,LugRomRos2019,KimKwaCho2022,KanGerMez2023}. Caching, the process of storing frequently accessed data closer to end-users, has long been vouched for in terrestrial networks (TN) to enhance content delivery speeds and reduce bandwidth requirements \cite{BasBenDeb2014,LiuCheYan2016,VuChaOtt2018}. However, in regions where optical fiber links are absent or inadequate, LEO satellites offer an alternative solution for establishing a global network fabric \cite{SedFelLin2020,VuPoiCha2020,LinLinCo2021}. 
By strategically deploying caching servers within these networks, content providers can mitigate latency, bandwidth limitations, and congestion issues that typically plague remote areas, ensuring faster access to frequently requested content. 
The use of NTN for caching not only addresses the connectivity gaps but also has the potential to revolutionize content delivery by extending the reach of digital services to previously untapped populations.

\blue{A major evolution of 5G and beyond is the integration of TN and NTN~\cite{3GPP38811,3GPP38821}.
Beyond merely filling connectivity gaps, NTN is valued for appropriately complementing TN performance shortfalls, thereby enhancing the overall efficiency of integrated systems~\cite{GerLopBen2022}.
This poses new challenges for efficient NTN content placement in the combined TN-NTN environment.
Due to large-scale and heterogeneous system topology, NTN and TN need to supplement each other to manage content delivery effectively.
Moreover, the dynamic nature of NTN, driven by satellite movement, necessitates a flexible content delivery strategy that can adapt to varying NTN capabilities.
Therefore, developing an effective and reliable NTN edge content broadcast strategy is essential to ensure efficient content delivery in integrated TN and NTN systems.}

\blue{Recent work has explored the use of NTN in content delivery, particularly focusing on leveraging LEO satellites for efficient cache content broadcasting in 5G edge wireless networks~\cite{DowFraSha2020,HeZhoWu2021,JiWuJiang2020,ZhuJiaKua2022}.
Existing studies have investigated NTN beam enhancement strategies to improve satellite transmission rates~\cite{LinNiuAn2022,LinLinCha2021}, and have positioned the role of NTN as traffic offloading for TN, developing algorithms to maximize NTN delivery throughput~\cite{HeLiWan2024,DenDiChe2020}.}
\blue{Additionally, several works have considered TN and NTN in an integrated manner, proposing radio resource allocation strategies to enhance the overall transmission capability of the integrated system~\cite{HanLiaPen2022,LiXueWei2020,JiaSheLi2022,CheMenHan2022}.}
Although these contributions have helped in understanding the potential role of NTN in delivering wireless edge caching content, they did not focus on the impact of LEO satellite mobility, producing a highly dynamic coverage pattern on the ground \cite{WanJiaWan2023,WanLiChe2022,JiZhoShe2022}. 
\blue{Moreover, existing approaches tend to sequentially optimize NTN delivery for each moment, resulting in suboptimal global content placement efficiency by precluding placement of files at later times when NTN coverage and broadcast capability may be more favorable, based on predictable satellite movement.
}
Indeed, optimal wireless content placement must consider both satellite mobility and the distribution of content popularity across the satellites' footprint.


\subsection{Contribution and Summary of Results}

This paper introduces a novel approach for optimizing wireless edge cache content placement using NTN. 
\blue{While existing methods typically focus either on enhancing network throughput at the satellite transmission side or maximizing cumulative user reception rates at the receiving side, our approach considers both sides and dynamically selects content for placement via NTN links, taking into consideration both the popularity distribution of the content and its suitability for delivery through NTN links. 
Unlike previous studies that have largely overlooked performance comparisons between NTN and TN at different times, our method treats NTN as a competitor to TN.
The suitability factor also accounts for the predictable orbital motion of NTN base stations (BSs) and for their capability of ensuring efficient content delivery to a certain geographical area.
}
The main contributions of our work can be summarized as follows:
\begin{itemize} 
\item 
We formulate the cache placement optimization problem within an integrated TN-NTN system to minimize the content delivery time. To address this problem, we propose heuristic approaches that determine the content to be placed via NTN links from the content's popularity distribution and the varying NTN coverage pattern.
\item 
We carry out system-level case studies based on a practical LEO constellation to evaluate the performance of our proposed approaches in various scenarios. The results demonstrate that our methods significantly improve the placement speed compared to state-of-the-art alternatives that overlook network mobility.
\item 
We demonstrate that the advantages offered by NTN links over standalone wireless TN solutions are particularly prominent in the initial stages of content delivery, when the most popular content is placed. We also confirm that the higher the correlation of content popularity across geographical regions, the more significant the advantage of NTN-based broadcast delivery becomes.
\end{itemize}

The rest of this paper is organized as follows. Section~II details our modeling assumptions for the network topology, wireless propagation channel, and content popularity and placement. In Section~III, we optimize NTN content placement iteratively on a time slot basis, whereas in Section~IV we tackle a joint optimization across time slots to further reduce the placement time. Section~V presents system-level case studies to evaluate the performance of
our proposed approaches in various scenarios. Section~VI concludes the article.
\section{System Model}

In this section, we introduce the following: (a) the network topology under consideration, (b) the wireless propagation channel model for terrestrial and non-terrestrial links, (c) the content file popularity model, and (d) the content placement model. The main notations employed are provided in Table~\ref{tab:tableNotation}.

\begin{figure}[!t]
\centering
\includegraphics[width=\figwidth]{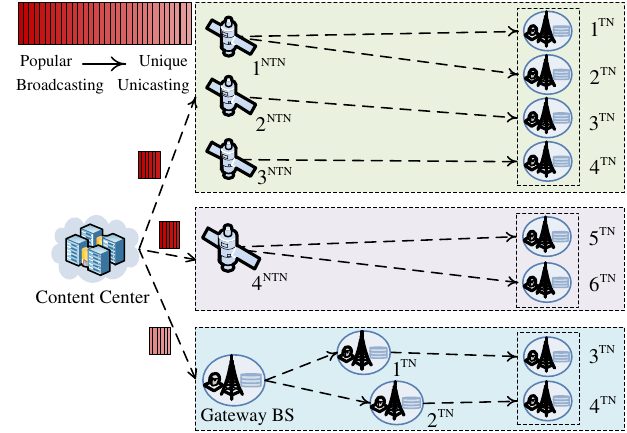}
\caption{Illustration of cache content placement in an integrated TN-NTN. More/less popular content is respectively broadcast/unicast via NTN/TN links.}
\label{fig:systemModel}
\vspace{-1em}
\end{figure}

\begin{table}[!t]
\renewcommand{\arraystretch}{1.3} 
\caption{Main notations used throughout the paper \label{tab:tableNotation} }
\centering
\begin{tabular}{c|c}

\toprule
\multicolumn{2}{c}{\textbf{Network topology}}\\ 
\hline 
$r$ & Geographical regions \\ 
\hline
$b^{\textrm{TN}}$, $b^{\textrm{NTN}}$ & TN and NTN BSs, $b^{\textrm{TN}} \in \ncalB^{\textrm{TN}}$ and $b^{\textrm{NTN}} \in \ncalB^{\textrm{NTN}}$ \\ 
\hline 
$\mathds{1}_{i,j}^{\textrm{est}}$ & NTN link between BSs $i^{\textrm{NTN}}$ and  $j^{\textrm{TN}}$ \\ 
\hline 
\multicolumn{2}{c}{\textbf{Propagation channel}}\\ 
\hline 
$P_i$ & Transmission power of BS $i$\\ 
\hline 
$G_{i,j}$ & Large-scale power gain between BSs $i$ and $j$\\ 
\hline 
$h_{i,j}$ & Small-scale block fading between BSs $i$ and $j$\\ 
\hline 
$\snr_{i,j}$, $\mathcal{R}_{i,j}$ & SNR and achievable rate between BSs $i$ and $j$ \\ 
\hline
\multicolumn{2}{c}{\textbf{Content popularity model}}\\ 
\hline 
$f$ & File with size $s_f$, $f=1,2, \ldots,N_{\mathrm{f}}$ \\ 
\hline
$p_{f,r}$ & Popularity of $f$ in region $r$ \\ 
\hline
$\alpha_{r}$ & Skewness of the popularity distribution in region $r$ \\ 
\hline
$\rho$ &  Coefficient of cache file correlation across regions \\
\hline
$\mathbf{p}_{r}$  &  File popularity in region $r$\\
\hline
\multicolumn{2}{c}{\textbf{Content placement model}}\\ 
\hline 
$t$, $T_t$ & NTN service time slot and duration, $t=1, \ldots, N_{\mathrm{t}}$ \\ 
\hline 
$\mathds{1}_{f,b}^{\textrm{ass}}$& Assignment of file $f$ to BS $b^{\textrm{TN}}$ for caching \\
\hline
$\mathds{1}_{f,t}^{\textrm{pla}}$ & Placement of file $f$ by NTN in $t$ \\
\hline
$\mathcal{R}_{b}^{\textrm{TN}}$, $\mathcal{R}_{b}^{\textrm{NTN}}$ & Delivery data rate to $b^{\textrm{TN}}$ via TN  and via NTN \\
\hline
$\tau_{f}^{\textrm{TN}}$,  $\tau_{f,t}^{\textrm{NTN}}$ & Delivery time for file $f$ via TN and via NTN \\ 
\hline
\multicolumn{2}{c}{\textbf{Content placement optimization}}\\ 
\hline 
$d_{b}^{\textrm{TN}}$ & Number of TN transmission hops to $b^{\textrm{TN}}$ \\ \hline
$\mu_{f,t}^{\textrm{part}}$, $\mu_{f,t}^{\textrm{sup}}$ & Indicators of NTN participation 
and NTN superiority  \\ \hline
$\mu_{f,t}$ & NTN delivery advantages metric \\ \hline
$\mathcal{F}_t$  & Files to be placed in time slot $t$ \\ \hline
$\ncalG =(V,  E,  W )$ & NTN placement graph (NPG) \\ \hline
$v$, $e_{i,j}$, $\omega_{i,j}$ & Vertex in $V$, edge in $E$, weight in $W$ in an NPG \\ \hline
$(f_v,t_v) $ & Coordinates for vertex $v$, $f_v \in \{f\}$, $t_v \in \{t\}$\\ \hline
$\ncalP $ & A path in $\ncalG$ combined with $\{e_{i,j}\}$ \\ \hline

\bottomrule
\end{tabular}
\vspace{-1em}
\end{table}


\subsection{Network Topology}

The network under consideration, illustrated in Fig.~\ref{fig:systemModel}, comprises the following elements, listed from right to left:
\begin{itemize}
\item A terrestrial network (TN) with edge base stations (BSs) $b^{\textrm{TN}} \in \ncalB^{\textrm{TN}}$,  
where all BSs within a certain region $r$ are connected to each other via microwave links forming a tree topology. 
\blue{Each region comprises a gateway BS, directly connected to the content center via optical fiber links and serving as the root node for intra-regional content transmission.}
\item A non-terrestrial network (NTN) consisting of a constellation of low Earth orbit (LEO) satellite BSs $b^{\textrm{NTN}} \in \ncalB^{\textrm{NTN}}$. 
Each satellite may receive files from the content center and broadcast them to some of the edge TN BSs.
\item 
\blue{A content center, located outside of each region and directly connected to TN and NTN gateways, holding files that can be fetched by TN edge BSs for caching.}
Its role is to determine whether each file should be delivered to TN edge BSs through the TN or NTN segments. The content center oversees a geographical area partitioned into regions $r$.
\end{itemize}

In the sequel, we therefore consider two types of links:
\begin{itemize}
\item \emph{TN links}, connecting two TN BSs $i^{\textrm{TN}}$ and $j^{\textrm{TN}}$.
\item \emph{NTN links}, between a NTN BS $i^{\textrm{NTN}}$ and a TN BS $j^{\textrm{TN}}$.
\end{itemize}
For TN links, we assume each BS to be able to connect to neighboring BSs within a certain coverage distance, so that a TN BS can receive files from the regional gateway via multiple hops. 
NTN links are affected by the LEO satellite elevation angle, whereby angles closer to $90^{\circ}$ yield shorter LEO-to-BS distances and are more likely to be in line-of-sight (LoS), 
\blue{whereas angles below $10^{\circ}$ are unable to ensure effective TN BS access and reliable data transmission~\cite{3GPP38811,3GPP38821}.}
\blue{When multiple TN BSs near region edges require the same content, the TN delivery speed is significantly reduced due to shared gateway transmission capacity, making NTN broadcasting a more effective solution.} 
For convenience, we define an indicator function $\mathds{1}_{i,j}^{\textrm{est}}$, whose value is one when an NTN link between BSs $i^{\textrm{NTN}}$ and $j^{\textrm{TN}}$ is established, and zero otherwise. 
\blue{We assume each TN BS to establish a link with at most one NTN BS at a time and follows a location-based handover strategy that always maintains NTN connectivity with the nearest satellite at any given time.}


\subsection{Propagation Channel Model}

The following models the channel and achievable rates. Further details are provided in Section~V and Table~\ref{tab:tableparametersNTNTNContentPopularity}.

\subsubsection*{Channel model}

All radio links are affected by path loss and lognormal shadow fading, both dependent on the link LoS condition
and modeled as in~\cite{3GPP38901,3GPP38811,3GPP38821}. Compared to a TN link, the signal travelling on an NTN link undergoes several extra stages of propagation. 
As a result, the total NTN path loss consists of additional terms accounting for the attenuation due to atmospheric gases and scintillation \cite{ITU}. 
We assume the LEO satellite antenna generating seven beams to be a typical reflector with circular aperture, and each TN BS to be equipped with a very small aperture terminal (VSAT). 
We denote $G_{i,j}$ the total large-scale gain between any two BSs $i$ and $j$, comprising path loss, shadow fading, and antenna gain at both ends. 
Similarly, we denote $h_{i,j}$ the small-scale block fading between the two BSs. 

\subsubsection*{Achievable data rates} 

The signal-to-noise ratio (SNR) on the link connecting BSs $i$ and $j$ is given by
\begin{equation}
  \vspace{-0.2cm}
  \snr_{i,j} = \frac{P_i \cdot G_{i,j} \cdot |h_{i,j}|^2 }{\sigma^2_j},
  \label{SNR_TN}
\end{equation}
where $P_i$ is the transmit power of BS $i$ and $\sigma^2_j$ is the thermal noise variance at BS $j$. Assuming that all edge BSs are distributed sparsely enough for the link to be interference-free, the corresponding achievable rate $\mathcal{R}_{i,j}$ can be obtained as
\begin{equation}
    \mathcal{R}_{i,j} = B_{i,j} \cdot \mathbb{E} \left[ \log_2 (1 + \snr_{i,j}) \right],
    \label{rates_TN}
\end{equation}
with $B_{i,j}$ denoting the bandwidth allocated to the link and where the expectation is taken over the small-scale fading.

\blue{For NTN links, the rapid movement of LEO satellites leads to variations in path loss which mainly depends on the transmission distance and in turn is a function of satellite altitude and elevation angle.
Shadow fading is modeled as a random variable following a normal distribution, with variance also influenced by the elevation angle, as detailed in~\cite{3GPP38811}.
Consequently, both $G_{i,j}$ and $h_{i,j}$ (and in turn, $\snr_{i,j}$ and $\mathcal{R}_{i,j}$) depend on the distance and elevation angle between BSs $i^{\textrm{NTN}}$ and $j^{\textrm{TN}}$. 
Their values vary with time and can be predicted from the satellite orbital information.}
At any given time, we define the NTN link duration $T_{i,j}$ between BSs ${i}^{\textrm{NTN}}$ and $j^{\textrm{TN}}$ as the remaining time until the condition $\mathds{1}_{i,j}^{\textrm{est}}$ changes its value from 1 to 0. 
\blue{For any TN BS $i$, we can determine its NTN connectivity as a LEO satellite sequence, given the handover strategy and predictable satellite movement.
Then, We can partition the time axis into slots $t=1, \ldots, N_{\mathrm{t}}$, such that all established NTN links (i.e., those with $\mathrm{1}_{i,j}^{\textrm{est}}=1$) remain as such within a slot, and at least one changes its value to zero across slots, thereby forming a sequence of $N_{\mathrm{t}}$ static NTN connectivity snapshots.}
At any given time, the remaining duration $T_t$ of the current slot $t$ can therefore be computed as 
\begin{equation} \label{GrindEQ__18_} 
T_t \,= \min\limits_{i, j \,|\,  \mathds{1}_{i,j}^{\textrm{est}}=1} \left\{T_{i,j} \right\}.
\end{equation} 



\subsection{Content Popularity Model}

We now introduce the cache content popularity model within each individual region and detail how content popularity is correlated across regions.

\subsubsection*{Content popularity within a certain region} 
The content file popularity is assumed to obey the Zipf distribution \cite{LiZhaLi2020}. We express the popularity $p_{f,r}$ of file $f$ in region $r$ as
\begin{equation}
p_{f,r} = \left[{f^{\alpha_{r}} \cdot \sum _{g=1}^{N_{\mathrm{f}}} g^{-\alpha_{r}} }\right]^{-1}, 
\label{GrindEQ__1_} 
\end{equation} 
where $f=1,2,\ldots,N_{\mathrm{f}}$, $\alpha_{r} \in \mathbb{R}^+$ is a skewness factor that controls the content popularity within $r$. Based on \eqref{GrindEQ__1_}, we first calculate the file popularity in the first region  and obtain a ranking list $p_{1,1} \ge p_{2,1} \ge \ldots \ge p_{N_{\mathrm{f}},1}$, with $\sum_{k=1}^{N_{\mathrm{f}}} p_{k,1} = 1$.
Let $\ncalB_r^{\textrm{TN}}\subseteq\ncalB^{\textrm{TN}}$ be the set of TN BSs in $r$ and let $|\ncalB_r^{\textrm{TN}}|$ denote its cardinality. The expected number of TN BSs requesting file $f$ in region $r$ is then given by $|\ncalB_r^{\textrm{TN}}| \cdot p_{f,r}$.

\subsubsection*{Content correlation across regions} 
 We assume the file popularity to vary across regions, with a coefficient $\rho $ modeling its correlation. Without loss of generality, we map each region $r$ to geographical coordinates $(\mathrm{x}_r,\mathrm{y}_r)\in \left\{\mathbb{N}^+\right\}^2$. 
After computing the file popularity list $\mathbf{p}_{1} = \left[p_{1,1}, \ldots, p_{N_{\mathrm{f}},1} \right]$ for the first region in $(1,1)$, the file popularity list $\mathbf{p}_{r}$ for region $r$ is modeled as
\begin{equation} \label{GrindEQ__2_} 
\begin{aligned}
& \mathbf{p}_r= \left\{ \begin{matrix}
	\rho \, \mathbf{p}_{\mathrm{x}_r-1,1}+\sqrt{1-\rho ^2} \, \mathbf{p}_{\mathrm{x}_r,1}&	\mathrm{x}_r>1, \mathrm{y}_r=1 \vspace*{0.2cm}\\
	\rho \, \mathbf{p}_{1,\mathrm{y}_r-1} + \sqrt{1-\rho ^2} \, \mathbf{p}_{1,\mathrm{y}_r}&		\mathrm{x}_r=1, \mathrm{y}_r>1 \vspace*{0.2cm}\\
    \begin{aligned}
        & \quad\quad \rho \, \left( \mathbf{p}_{\mathrm{x}_r,\mathrm{y}_r-1} + \mathbf{p}_{\mathrm{x}_r-1,\mathrm{y}_r} \right)
        \rule[-0.3cm]{0pt}{0.3cm}\\
        &-\rho^2 \, \mathbf{p}_{\mathrm{x}_r-1,\mathrm{y}_r-1}+\left( 1-\rho ^2 \right) \mathbf{p}_{\mathrm{x}_r,\mathrm{y}_r}
    \end{aligned}
    & 		\mathrm{x}_r>1, \mathrm{y}_r>1\\
\end{matrix} \right.
\end{aligned}
\end{equation}
where we can derive $\rho^{|\mathrm{x}_r - \mathrm{x}_{r'} | + |\mathrm{y}_r - \mathrm{y}_{r'} |}$ as the file popularity correlation between regions $r$ and $r'$. The values of $\{\alpha_{r}\}$ and $\rho$ thus allow to model the content popularity across the whole geographical area.


\subsection{Content Placement Model}

We define a binary indicator $\mathds{1}_{f,b}^{\textrm{ass}}$ whose value is one if file $f$ is assigned to BS $b^{\textrm{TN}}$ for caching and zero otherwise. Since each TN BS has a finite cache size and the file popularity is probabilistic, we assume that the content center plans for only $\lfloor |\ncalB_r^{\textrm{TN}}| \cdot p_{f,r} \rfloor $ BSs in region $r$ to cache file $f$ in decreasing order of popularity $p_{f,r}$ until the cache of all BSs in $r$ reaches saturation. Such assignments are recorded through $\{\mathds{1}_{f,b}^{\textrm{ass}}\}$. 

\blue{We assume that content placement takes place sequentially in each segment, with each file being assigned for delivery to the edge BSs through either broadcast (via NTN links) or unicast (via multi-hop TN links).} 
In each time slot $t$, the content center determines which files to be broadcast over NTN links. We denote each assignment with a binary indicator $\mathds{1}_{f,t}^{\textrm{pla}}$, whose value is one if file $f$ is allocated for NTN delivery in time slot $t$ and zero otherwise. 

\subsubsection*{Content placement via TN links}

\blue{When using TN links, a file $f$ is delivered to a BS $b^{\textrm{TN}}$ in a multi-hop fashion with store-and-forward relay protocol, with $d_{b}^{\textrm{TN}}$ denoting the distance between the gateway and $b^{\textrm{TN}}$ expressed in number of transmission hops required.}
The rate $\mathcal{R}_{b}^{\textrm{TN}}$ between the gateway and $b^{\textrm{TN}}$ is given by the lowest rate on any of the hops, each of the latter governed by (\ref{rates_TN}). 
The placement time $\tau_{f}^{\textrm{TN}}$ of file $f$ for the served area is then the maximum across all BSs requiring such file, i.e., 
\begin{equation}
\tau_{f}^{\textrm{TN}}=\max\limits_{b^{\textrm{TN}} |\, \mathds{1}_{f,b}^{\textrm{ass}}=1} \left\{s_f/\mathcal{R}_{b}^{\textrm{TN}} \right\},
\label{TNrate} 
\end{equation}
where $s_f$ denotes the size of file $f$.

\subsubsection*{Content placement via NTN links} 

As the NTN topology remains unchanged in each time slot, we approximate the data rate of an NTN link with its average value. By neglecting the delay incurred on the high-capacity optical link that connects the content center to the NTN BS, we then compute $\tau_{f,t}^{\textrm{NTN}}$ for a file $f$ in slot $t$ as the transmission time over the NTN link,
\begin{equation}
\tau_{f,t}^{\textrm{NTN}}=\max\limits_{b^{\textrm{TN}} |\, \mathds{1}_{f,b}^{\textrm{ass}}=1} \left\{s_f/\mathcal{R}_{b}^{\textrm{NTN}}\right\},
\label{NTNrate} 
\end{equation}
where 
$\mathcal{R}_{b}^{\textrm{NTN}}$ denotes the rate on the link between BSs $i^{\textrm{NTN}}$ (delivering the file) and $b^{\textrm{TN}}$ (receiving it). While not indicated explicitly for ease of notation, the rate $\mathcal{R}_{b}^{\textrm{NTN}}$ varies across time slots and it can be computed via (\ref{rates_TN}).
\blue{Hence, $\tau_{f,t}^{\textrm{NTN}}$ varies with satellite mobility and content assignments, guided by the file placement indicator $\mathds{1}_{f,t}^{\textrm{pla}}$ to determines whether file $f$ is delivered within time slot $t$ via NTN.}

In what follows, we first optimize NTN content placement iteratively on a time slot basis (Section~III), and then we tackle a joint optimization across time slots to further reduce the content placement time (Section~IV).
\section{Sequential Content Placement Optimization}

We begin by addressing content placement optimization through a heuristic strategy that determines the values $\mathds{1}_{f,t}^{\textrm{pla}}$ on a slot-by-slot basis, i.e., separately for each time slot $t$. 

\subsection{Problem Formulation}

At each time slot, $t$, the content center aims to maximize the total delivery time savings. This is achieved by opportunistically delivering files via either TN or NTN links.
\begin{Problem} \label{P1}
Sequential content placement optimization
\begin{align} \label{GrindEQ__25_} 
\max\limits_{\mathds{1}_{f,t}^{\mathrm{pla}}} & \,\sum_{f=1}^{ N_\mathrm{f}} \,\mathds{1}_{f,t}^{\mathrm{pla}} \,\cdot \,\left(\tau_{f}^{\mathrm{TN}} - \tau_{f,t}^{\mathrm{NTN}}\right)\\
\mathrm{s.t.}\; 
& \,\sum\limits_{f=1}^{ N_\mathrm{f}} \,\mathds{1}_{f,t}^{\mathrm{pla}} \cdot \,\tau_{f,t}^{\mathrm{NTN}} \le T_{t}
\tag{\ref{GrindEQ__25_}a}\\
& \,\sum_{t'=1}^{t} \,\mathds{1}_{f,t'}^{\mathrm{pla}}  \le 1,\; \enspace f =1,\ldots,N_{\mathrm{f}} \tag{\ref{GrindEQ__25_}b}.
\end{align} 
\end{Problem}
\noindent 
\blue{In Problem~1, (\ref{GrindEQ__25_}) represents the time difference between TN and NTN delivery, effectively capturing the total amount of time saved by placing certain files $f$ in the current time slot $t$ via the NTN segment rather than the TN.}
The constraint (\ref{GrindEQ__25_}a) ensures that the delivery time via NTN links within the current slot does not exceed the duration of the slot itself, whereas (\ref{GrindEQ__25_}b) ensures any file that has been placed in a previous slot is not placed again in the current slot. 
Given the delivery times $\tau_{f}^{\mathrm{TN}}$ and $\tau_{f,t}^{\mathrm{NTN}}$ of each file $f$ via TN and NTN links, solving Problem~1 at time slot $t$ entails determining the values $\{ \mathds{1}_{f,t}^{\textrm{pla}}\}$, i.e., determining for the current time slot $t$ which files to be placed via the NTN segment.

\begin{Remark}
For given $t$, $\tau_f^{\mathrm{TN}}$, and $\tau_{f,t}^{\mathrm{NTN}}$, (\ref{GrindEQ__25_}) has a finite solution space and is polynomial-time verifiable since the size of every feasible set $\{ \mathds{1}_{f,t}^{\mathrm{pla}}\}$ is polynomially bounded by the number of cache files to be placed, the latter being finite. 
If one regards $T_{t} $ as the knapsack capacity and $\tau_f^{\mathrm{TN}} - \tau_{f,t}^{\mathrm{NTN}}$ as the price of item $f$, Problem~1 is equivalent to a knapsack problem that seeks to maximize the total price under a limited capacity, and is therefore NP-complete. As a result, solving Problem 1 through traversal search incurs a complexity in the order of $O(2^{N_\mathrm{f}})$ and is not a viable approach.
\end{Remark}

\subsection{Proposed Sequential Content Placement Approach}

We now propose a heuristic approach to maximize the total delivery time reduction in  (\ref{GrindEQ__25_}). This approach opportunistically leverages NTN links and it is based on two metrics, introduced as follows, that capture the advantage of NTN-based broadcast delivery.

\subsubsection*{NTN participation indicator $\mu^{\textrm{part}}$} 

We recall the indicators $\mathds{1}_{i,j}^{\textrm{est}}$ and $\mathds{1}_{f,j}^{\textrm{ass}}$, whose value is one respectively if a link between BSs $i^{\textrm{NTN}}$ and $j^{\textrm{TN}}$ is established and if BS $j^{\textrm{TN}}$ is assigned file $f$. We can then employ $\mathds{1}_{f,j}^{\textrm{ass}}\cdot \mathds{1}_{i,j}^{\textrm{est}}$ to indicate whether BS $j^{\textrm{TN}}$ should receive file $f$ and has a link with BS $i^{\textrm{NTN}}$. The latter allows us to define an \emph{NTN participation indicator} $\mu_{f,t}^{\textrm{part}} \in [0,1]$ as the fraction of NTN BSs suitable to deliver file $f$ out of all NTN BSs providing service in slot $t$, given by
\begin{equation} \label{NTNpart} 
\mu_{f,t}^{\textrm{part}} = \frac{\sum_{i\in \ncalB^{\textrm{NTN}}} \left(1 - \left(\prod_{j\in\ncalB^{\textrm{TN}}}  \left(1 -  \mathds{1}_{f,j}^{\textrm{ass}}\cdot \mathds{1}_{i,j}^{\textrm{est}}  \right)\right) \right)}
{\sum_{i\in \ncalB^{\textrm{NTN}}} \left(1 - \left(\prod_{j\in\ncalB^{\textrm{TN}}}  \left(1 - \mathds{1}_{i,j}^{\textrm{est}}\right)\right) \right) } .
\end{equation} 
The value of $\mu_{f,t}^{\textrm{part}}$ can be calculated in each time slot $t$ to infer the degree of participation of NTN BSs in the delivery of file $f$. Intuitively, higher values of $\mu_{f,t}^{\textrm{part}}$ indicate a higher suitability for file $f$ to be deployed via NTN broadcast.

\subsubsection*{NTN superiority indicator $\mu^{\textrm{sup}}$}

We recall that $d_{b}^{\textrm{TN}}$ denotes the number of hops required to deliver content to TN BS $b^{\textrm{TN}}$ via TN links. Then the average number of TN hops $d_{f,t}^{\textrm{ave}}$ that can be avoided by instead delivering file $f$ via NTN links in slot $t$ can be calculated as
\begin{equation} \label{NTNhop} 
d_{f,t}^{\textrm{ave}} \!=\! \frac{\sum\limits_{i\in\ncalB^{\textrm{NTN}}} \!\!\frac{
\sum\limits_{j\in\ncalB^{\textrm{TN}}} (\mathds{1}_{f,j}^{\textrm{ass}}\cdot \mathds{1}_{i,j}^{\textrm{est}}\cdot d_{j}^{\textrm{TN}} ) }{
\sum\limits_{j\in\ncalB^{\textrm{TN}}}  \mathds{1}_{f,j}^{\textrm{ass}}\cdot \mathds{1}_{i,j}^{\textrm{est}} 
}
\!\left(\!1 \!-\! \left(\prod\limits_{j\in\ncalB^{\textrm{TN}}}  \left(1 \!-\! \mathds{1}_{f,j}^{\textrm{ass}}\cdot \mathds{1}_{i,j}^{\textrm{est}}\right)\right) \right)}
{\sum\limits_{i\in\ncalB^{\textrm{NTN}}}\left(1 - \left(\prod\limits_{j\in\ncalB^{\textrm{TN}}}  \left(1 - \mathds{1}_{f,j}^{\textrm{ass}}\cdot \mathds{1}_{i,j}^{\textrm{est}}\!\right)\!\right) \!\right)}
\end{equation} 
and it can be normalized to an \emph{NTN superiority indicator} $\mu_{f,t}^{\textrm{sup}}\in[0,1]$ through the following rescaling
\begin{equation} \label{NTNsup} 
\mu_{f,t}^{\textrm{sup}} = \frac{d_{f,t}^{\textrm{ave}}-\min_f \{d_{f,t}^{\textrm{ave}}\}}{\max_f \{d_{f,t}^{\textrm{ave}}\}-\min_f \{d_{f,t}^{\textrm{ave}}\}}.  
\end{equation} 
Intuitively, higher values of $\mu_{f,t}^{\textrm{sup}}$ indicate a higher suitability for a file $f$ to be delivered via NTN links, since delivering the same file via TN would entail a larger number of transmission hops, hence a longer delivery time.

\subsubsection*{Overall NTN suitability indicator $\mu$}

We finally combine indicators $\mu_{f,t}^{\textrm{part}}$ and $\mu_{f,t}^{\textrm{sup}}$ into a single metric $\mu_{f,t}\in[0,1]$, capturing the overall suitability of file $f$ to be delivered via NTN links in time slot $t$ and defined as follows 
\begin{equation} 
\mu_{f,t}=\frac{\left[\mu_{f,t}^{\textrm{part}} \right]^{\beta } \cdot \left[\mu_{f,t}^{\textrm{sup}}\right]^{1-\beta } }{\sum\limits_{k=1}^{N_{\mathrm{f}}}\left\{ \left[\mu_{k,t}^{\textrm{part}} \right]^{\beta } \cdot \left[\mu_{k,t}^{\textrm{sup}}\right]^{1-\beta } \right\}  }, \label{NTNpop} 
\end{equation} 
where the coefficient $\beta$ is introduced to trade off the participation and superiority criteria in (\ref{NTNpart}) and (\ref{NTNsup}). 

\subsubsection*{Content placement algorithm}

\begin{figure}[!t]
\centering
\includegraphics[width=\figwidth]{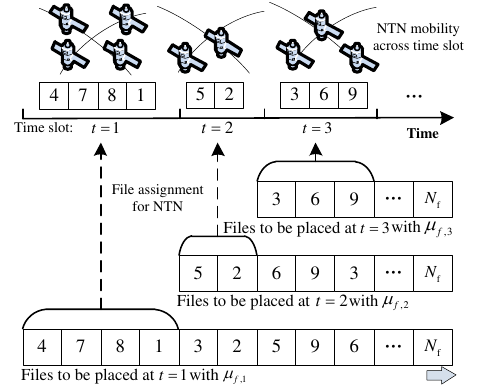}
\caption{Illustration of the proposed sequential file assignment approach.}
\label{fig:SFA}
\end{figure}

Let $\mathcal{F}_t$ denote the set of files to be placed in a given time slot $t$. The content center then ranks all files $f \in \mathcal{F}_{t}$ according to the indicator $\mu_{f,t}$ in (\ref{NTNpop}) and deploys as many files as possible within $t$ via the NTN in decreasing order of $\mu_{f,t}$. Formally, such decision corresponds to setting $\mathds{1}_{f,t}^{\textrm{pla}}$ to one when file $f$ is allocated for NTN placement in time slot $t$. If condition (\ref{GrindEQ__25_}a) is no longer met during this process, the content center stops allocating files for NTN placement as the number of files that can be placed within time slot $t$ has reached saturation. 
After slot $t$, the set of files yet to be delivered is calculated as $\mathcal{F}_{t+1}=\mathcal{F}_{t}-\{f \,|\, \mathds{1}_{f,t}^{\textrm{pla}} = 1\}$ and the new values $\mu_{f,t+1}$ are calculated for all files in $\mathcal{F}_{t+1}$.
The procedure is repeated until the last file has been deployed, i.e., until $\mathcal{F}_t = \emptyset$. 
The proposed approach is denoted NTN \emph{sequential file assignment (SFA)} and it is illustrated in Fig.~\ref{fig:SFA} and detailed in Algorithm \ref{alg1}.

\begin{algorithm}[!t] 
\caption{Proposed sequential file assignment (SFA).} 
\label{alg1} 
\begin{algorithmic}[1] 
\REQUIRE ~~\\ 
BSs $\ncalB^{\textrm{NTN}}$ and $\ncalB^{\textrm{TN}}$, files $\mathcal{F}_1$, popularity $\{\mathds{1}_{f,b}^{\textrm{ass}}\}$, $t=1$; \\
\ENSURE ~~\\ 
NTN file assignments $\{\mathds{1}_{f,t}^{\textrm{pla}}\}$;
\WHILE{$\mathcal{F}_t \neq \emptyset$}
    \STATE Compute NTN link conditions $\mathds{1}_{i,j}^{\textrm{est}}$;
    \STATE Compute $\tau_{f}^{\textrm{TN}}$ and $\tau_{f,t}^{\textrm{NTN}}$ from (\ref{TNrate}) and (\ref{NTNrate}); 
    \FOR {$f\in \ncalF_t$}
        \STATE Compute $\mu_{f,t}^{\textrm{part}}$ from (\ref{NTNpart}) and $\mu_{f,t}^{\textrm{sup}}$ from (\ref{NTNsup});
        \STATE Compute $\mu_{f,t}$ from (\ref{NTNpop});   
    \ENDFOR
    \STATE Sort all files in $\ncalF_t$ in decreasing order of $\mu_{f,t}$;
    \WHILE{constraint (\ref{GrindEQ__25_}a) is met}
        \STATE Set $\mathds{1}_{f,t}^{\textrm{pla}} \gets 1$ for ordered files $f \in \ncalF_t$;
    \ENDWHILE
    \STATE Update $\mathcal{F}_{t+1} \gets \mathcal{F}_{t}-\{f \,|\, \mathds{1}_{f,t}^{\textrm{pla}} = 1\}$;
    \STATE $t \gets t+1$;
\ENDWHILE
\end{algorithmic}
\end{algorithm}

\subsubsection*{Computational complexity of SFA}
The SFA approach ranks all files $f \in \mathcal{F}_{t}$  and allocates them for each time slot $t$ until all $N_\mathrm{f}$ files are assigned over $N_{\mathrm{t}}$ slots, resulting in a complexity in the order of $O( N_{\mathrm{t}} \cdot {N_\mathrm{f}}^{2})$.

\begin{Remark}
\blue{Optimizing NTN content placement on a sequential basis restricts the solution space with respect to a joint optimization across all time slots, considering predictable satellite orbital motion, and may thus achieve smaller delivery time savings.} 
Indeed, as expressed by constraint (\ref{GrindEQ__25_}b), placing a file $f$ in time slot $t$ precludes the possibility of placing $f$ in subsequent slots where the NTN coverage and transmission capability may be more favorable.
\end{Remark}

In light of Remark 2, in the next section we tackle a global NTN placement optimization that accounts for the interdependencies across all time slots.
\section{Global Content Placement Optimization}

We now address content placement optimization across all time slots. Our approach uses a graph-based solution, expanding the solution space of Algorithm~1.

\subsection{Problem Formulation}

The content center aims at maximizing the total savings achieved across all time slots in terms of content delivery time by opportunistically delivering files via either TN or NTN links, as follows.

\begin{Problem} \label{P2}
Global content placement optimization
\begin{align} \label{P2formulation} 
\max\limits_{\mathds{1}_{f,t}^{\mathrm{pla}}} & \,\sum_{t=1}^{N_\mathrm{t}} \,\sum_{f=1}^{ N_\mathrm{f}} \,\mathds{1}_{f,t}^{\mathrm{pla}} \,\cdot \,\left(\tau_{f}^{\mathrm{TN}} - \tau_{f,t}^{\mathrm{NTN}}\right)\\
\mathrm{s.t.}\; 
& \,\sum\limits_{f=1}^{ N_\mathrm{f}} \,\mathds{1}_{f,t}^{\mathrm{pla}} \cdot \,\tau_{f,t}^{\mathrm{NTN}} \le T_{t}, \; \enspace t =1,\ldots,{N_\mathrm{t}} \tag{\ref{P2formulation}a}\\
& \,\sum_{t=1}^{{N_\mathrm{t}}} \,\mathds{1}_{f,t}^{\mathrm{pla}}  \le 1,\; \enspace f =1,\ldots,N_{\mathrm{f}} \tag{\ref{P2formulation}b}.
\end{align} 
\end{Problem}

\noindent In Problem~2, (\ref{P2formulation}) represents the total amount of time saved by placing a file $f$ in time slot $t$ via the NTN segment rather than the TN. The constraint (\ref{P2formulation}a) ensures that the delivery time via NTN links within each slot does not exceed the duration of the slot itself, whereas (\ref{P2formulation}b) ensures that each file is not delivered more than once across all slots.

\begin{Remark}\label{P1andP2}
We observe that Problem 2 subsumes Problem 1 when considering $N_\mathrm{t}$ time slots. Consequently, the feasible region of Problem 2 encompasses that of Problem 1, implying that solving Problem 2 is computationally not easier than solving Problem 1.
In a similar way as detailed in Remark 1 for Problem 1, Problem 2 is also NP-complete, ruling out traversal search.
\end{Remark}

\begin{figure*}[!t]
\centering
\includegraphics[width=1.5\figwidth]{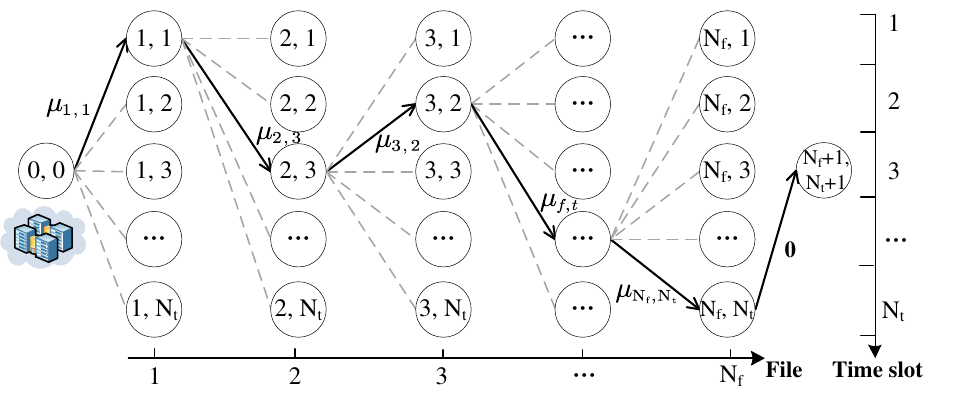}
\caption{Example of pathfinding within the NTN placement graph $\ncalG$ defined in (\ref{GrindEQ__36_}) and (\ref{GrindEQ__37_}) and constructed as detailed in Algorithm~\ref{alg2}.}
\label{fig:GFA}
\end{figure*}

\subsection{Proposed Global Content Placement Approach}

We propose a heuristic approach to maximize the total reduction in delivery time (\ref{P2formulation}). This approach opportunistically leverages NTN links and is based on similar metrics as those introduced in Section~III. However, for a global content placement optimization, the NTN delivery indicator $\mu_{f,t}$ in (\ref{NTNpop}) can be calculated in advance for each file $f$ for different time slots $t$, extending the scope of file selection in the time dimension. Considering all $N_{\mathrm{f}} \cdot N_{\mathrm{t}}$ values $\mu_{f,t}$ for all files $f$ and time slots $t$ allows to select $N_{\mathrm{f}}$ such values to maximize the total NTN delivery advantage in file placement. Our proposed solution to this problem entails the construction of a graph, as detailed in the sequel.

\subsubsection*{NTN placement graph model}

For a more intuitive problem formulation, we map all possible values of ${ \mu_{f,t}}$ into an NTN placement graph
\begin{equation} \label{GrindEQ__36_}
\ncalG =(V, \, E, \, W ), 
\end{equation}
where $V$, $E$, and $W$ are defined as follows:
\begin{itemize}
\item $V =\{ v \}$ is a set of $N_{\mathrm{f}} \cdot N_{\mathrm{t}}$ vertices mapped to coordinates $(f_v, t_v)$ and representing all NTN file assignment combinations. Specifically, a vertex $v$ with coordinate $(f_v, t_v)$ indicates the placement of file $f$ at time slot $t$ via the NTN segment.
\item $E =\{ e_{i,j} \} \subset \{V \times V \} $ is a set of edges, with $e_{i,j}$ pointing from vertex $i$ to vertex $j$.
\item $W =\{ \omega _{i,j} \}$ is a set of weights, with $\omega _{i,j} $ denoting the weight associated to edge $e_{i,j}$.
\end{itemize}

\noindent The NTN placement graph $\ncalG$ is illustrated in Fig.~\ref{fig:GFA} and it abides the the following conditions:
\begin{equation} \label{GrindEQ__37_} 
\left\{
\begin{array}{c} 
e_{i,j} \in E \quad \forall \, i,j \in V \enspace | \enspace f_j - f_i \geq 1 \vspace{2ex}\\ 
\omega_{i,j} =\mu_{f,t} \quad \forall e_{i,j} \in E \enspace | \enspace f_{j}=f, \enspace t_{j}=t.
\end{array}\right.
\end{equation} 
\noindent The first condition in (\ref{GrindEQ__37_}) ensures that the same file cannot be placed in different time slots, thus there cannot be an edge connecting two vertices corresponding to the same file. Moreover, the first condition ensures that any pair of vertices is connected by a single edge at most, from the lower to the higher file index, as exemplified in Fig.~\ref{fig:GFA}. 
The second condition in (\ref{GrindEQ__37_}) ensures that all edges pointing to the same vertex carry the same weight. Such weight is equal to $\mu_{f,t}$, the suitability metric associated to deploying file $f$ in slot $t$. 
The graph is completed by adding a source vertex $v_0=(0,0)$ and a destination vertex $v_{N_{\mathrm{f}} \cdot N_{\mathrm{t}}+1}=(N_{\mathrm{f}}+1, N_{\mathrm{t}}+1)$, the latter reached via a zero-weight edge. 
The process leading to the construction of graph $\ncalG$ is detailed in Algorithm \ref{alg2}. 

\begin{algorithm}[!t] 
\caption{Construction of the NTN placement graph.} 
\label{alg2} 
\begin{algorithmic}[1] 
\REQUIRE ~~\\ 
All files $\{f\}$ and time slots $\{t\}$;\\
\ENSURE ~~\\ 
NTN placement graph $\ncalG =(V, \, E, \, W )$;
\STATE Initialize $V = \emptyset$;
\FOR {$f=1$ to $N_{\mathrm{f}}$}
    \FOR {$t=1$ to $N_{\mathrm{t}}$}
        \STATE Add vertex $v = N_{\mathrm{t}}\cdot(f-1)+t$ to $V$ with coordinates $(f_v, t_v) = (f, t)$;
    \ENDFOR    
\ENDFOR
\STATE Add $v_0=(0,0)$, $v_{N_{\mathrm{f}} \cdot N_{\mathrm{t}}+1}=(N_{\mathrm{f}}+1, N_{\mathrm{t}}+1)$ to $V$;
\FOR {$i \in V$}
    \FOR {$j \in V$, $j \neq i$}
        \STATE Determine the existence of edge $e_{i,j} \in E$ from (\ref{GrindEQ__37_});
        \IF {$e_{i,j} \in E$}
            \STATE Compute weight $\omega_{i,j} \in W$ from (\ref{GrindEQ__37_}) and (\ref{NTNpop});
        \ENDIF
    \ENDFOR
\ENDFOR
\end{algorithmic}
\end{algorithm}


\begin{Proposition}\label{DAG}
Constraint (\ref{P2formulation}b) allows for an efficient representation of Problem 2 through a directed acyclic graph $\ncalG$, preserving the original solution space.
\end{Proposition}

\textit{Proof.}
Each of the $N_{\mathrm{f}} \cdot N_{\mathrm{t}}$ decision indicators $\mathds{1}_{f,t}^{\mathrm{pla}}$ in Problem 2 corresponds to a vertex in $\ncalG$. Since constraint (\ref{P2formulation}b) does not restrict deployment dependencies among different files $f$, we can arrange the vertices in $\ncalG$ in file index order, maintaining the solution space of Problem 2. Additionally, constraint (\ref{P2formulation}b) enforces that each file is assigned at most once, preventing edges in $\ncalG$ between vertices representing the same $f$ as well as edges directed towards vertices representing earlier files. Consequently, $\ncalG$ contains no strongly connected components, confirming it as a directed acyclic graph that effectively mirrors the solution space defined by (\ref{P2formulation}b).
$\hfill\blacksquare$

\subsubsection*{Global content placement strategy}

Once the NTN placement graph $\ncalG$ is constructed, moving through $\ncalG$ along the horizontal file axis and connecting $N_{\mathrm{f}}$ vertices corresponds to identifying an NTN content placement strategy. We therefore propose a heuristic approach to maximize the delivery time reduction in (\ref{P2formulation}) by finding a path $\ncalP$ in $\ncalG$ that yields the maximum sum value of the weights $\{ \mu_{f,t}\}$, as exemplified in Fig.~\ref{fig:GFA}. 
A path $\ncalP =\{ e_{i,j} \}$, embodying a feasible solution of Problem~2, is defined as a sequence of $|\ncalP | = N_{\mathrm{f}}+1$ consecutive edges $e_{i,j}$, connecting $v_0$ to $v_{N_{\mathrm{f}} \cdot N_{\mathrm{t}}+1}$ and thus selecting all $N_{\mathrm{f}}$ files. The pathfinding problem is formulated as follows.

\begin{Problem} \label{Problem3}
Maximum weight pathfinding in $\ncalG$
\begin{align} \label{P3}
\max\limits_{\ncalP} 
& \,\, \,\, \sum\limits_{e_{i,j} \in \ncalP} \, \mu_{i,j} \\  
\mathrm{s.t.}\; 
&\, \sum\limits_{j: \, e_{v_{0},j} \in \ncalP} \!\! e_{v_{0},j} = 1, \!\!  \sum\limits_{i: \, e_{i,v_{N_{\mathrm{f}} \cdot N_{\mathrm{t}}+1}} \in \ncalP} \!\!\!\! e_{i, {v_{N_{\mathrm{f}} \cdot N_{\mathrm{t}}+1}}} = 1 \tag{\ref{P3}a}\\
& \,\,\, \sum\limits_{i: \, e_{i,j} \in \ncalP} \, e_{i,j} =  \sum\limits_{k: \, e_{j,k} \in \ncalP} \, e_{j,k},  \,\, \,\,  \forall j \in V \tag{\ref{P3}b}\\
& \! \sum_{e_{i,j}\in \ncalP \,|\, t_j=t} \tau_{f_i,t}^{\mathrm{NTN}} \le T_{t}, \,\, \,\, t=1, \ldots, N_{\mathrm{t}} \tag{\ref{P3}c}
\end{align} 
\end{Problem}

\noindent where (\ref{P3}c) in Problem~3 ensures that during pathfinding, the sum of the placement times of all files to be delivered within each slot $t$ does not exceed the duration $T_t$ of the slot itself.

\subsubsection*{Pathfinding approach}
To identify the highest-weight path in $\ncalG$ that meets constraint (\ref{P3}c), we adapt a modified version of the Bellman-Ford algorithm~\cite{LiLiWan2020}. The process includes:
\begin{enumerate}
    \item Initializing all path weight sums ${\mathrm{sum}[v]}$ from the source vertex $v_0$ to other vertices in $\ncalG$ as negative infinity.
    \item Updating $\mathrm{sum}[j]$ for each edge $e_{i,j}$ in $\ncalG$. If $\mathrm{sum}[i] + \omega_{i,j} > \mathrm{sum}[j]$ and $f_j$, $t_j$ satisfy (\ref{P3}c), setting $\mathrm{sum}[j] = \mathrm{sum}[i] + \omega_{i,j}$.
    \item Repeating step 2 for $N_{\mathrm{f}} + 1$ iterations to finalize $\mathrm{sum}[v]$ values for all vertices, including the destination node $v_{N_{\mathrm{f}} \cdot N_{\mathrm{t}}+1}$.
\end{enumerate}
We track each vertex's last hop with $\mathrm{pre}[j]$ to reconstruct the maximum-weight path $\ncalP$ by backtracking from $v_{N_{\mathrm{f}} \cdot N_{\mathrm{t}}+1}$ to $v_0$. 
If the condition $|\ncalP | = N_{\mathrm{f}}+1$ is met, this indicates that the delivery of all files can be completed in $N_{\mathrm{t}}$ time slots.

\begin{Remark}\label{SFAandGFA}
When $\ncalG$ encompasses $N_{\mathrm{f}} \cdot N_{\mathrm{t}}+2$ vertices, there exists a path $\ncalP$ satisfying $|\ncalP | = N_{\mathrm{f}}+1$, i.e., all $N_{\mathrm{f}}$ files can be placed in $N_{\mathrm{t}}$ slots.
This is because any solution derived from Algorithm~\ref{alg1} is also a valid path in $\ncalG$. However, such solution comes from a limited pathfinding scope, as noted in Remark~\ref{P1andP2}. In contrast, the solution obtained from pathfinding in $\ncalG$ is the result of exploring the entire solution space for file placement, which may allow to find a path that meets the condition $|\ncalP | = N_{\mathrm{f}}+1$ in a smaller graph $\ncalG$, i.e., to complete the file delivery within $n_{\mathrm{t}} < N_{\mathrm{t}}$ slots.
\end{Remark}

In the following, we vary the size of graph $\ncalG$ until finding a minimum number of time slots, $n_{\mathrm{t}}$, necessary to complete the delivery of all files.

\subsubsection*{Binary search-based optimal placement time}

We employ an enhanced binary search method to efficiently determine the smallest number of time slots $n_{\mathrm{t}}$ required to deliver all $N_{\mathrm{f}}$ files. 
Specifically, we employ a test-and-double binary search strategy \cite{WanSouLi2023}, where at each iteration we test whether $t$ slots are sufficient to deliver all files by running the pathfinding algorithm. We start from $t=1$ and double its value at each iteration, until $t$ slots are sufficient to place all files, thereby indicating that the optimal placement time satisfies $n_{\mathrm{t}} \in [t/2, \  t]$. We then carry out further iterations to fine-search within the range $[t/2, \  t]$ and determine the exact value of $n_{\mathrm{t}}$. The path $\ncalP$ employing $n_{\mathrm{t}}$ slots is then regarded as the optimal NTN file assignment strategy. This proposed approach is denoted the \emph{global file assignment (GFA)} and it is detailed in Algorithm~\ref{alg3}.
\blue{It follows that the final obtained $n_{\mathrm{t}}$ represents the minimum total NTN broadcast time, as no smaller slot number $t$ than $n_{\mathrm{t}}$ can achieve $|\ncalP|=N_{\mathrm{f}}+1$, marking the completion of delivering all $N_{\mathrm{f}}$ files.}
While requiring a similar number of iterations, this approach is more efficient than a traditional binary search, \blue{starting from} $t=N_{\mathrm{t}}$ and halving until $t$ slots are sufficient to place all files. Indeed, the complexity of running the pathfinding algorithm is higher for a larger graph, as indicated in the following.

\begin{algorithm}[!t] 
\caption{Proposed global file assignment (GFA).}
\label{alg3} 
\begin{algorithmic}[1] 
\REQUIRE ~~\\ 
NTN placement graph $\ncalG =(V ,E ,W )$,\\file number $f=N_{\mathrm{f}}$, time slot $t=1$, path $\ncalP_{t}=\emptyset$;\\
\ENSURE ~~\\ 
Maximum-weight path $\ncalP$ employing fewest time slots;
\WHILE{{$|\ncalP_{t} | < N_{\mathrm{f}}+1 $}}
\STATE {$t \gets t \times 2$;}
\STATE {Vary $\ncalG \gets $ $\{v | t_v \leq t\}$, $\{e_{i,v}\}$, $\{\omega_{i,v}\}$;}
\FOR{$v \in V $}
    \STATE Set $\mathrm{sum}[v] \gets -\infty$, $\mathrm{pre}[v] \gets \mathrm{null}$;
\ENDFOR
\STATE Set $\mathrm{sum}[v_0]=0;$
\FOR{$k=1$ to $N_{\mathrm{f}} + 1 $}
    \FOR{$e_{i,j} \in E $}
        \IF{$\mathrm{sum}[i] \!+\!  \omega_{i,j} \!>\! \mathrm{sum}[j]$ and $f_j, t_j$ meet (\ref{P3}c)} 
            \STATE Update $\mathrm{sum}[j] \gets \mathrm{sum}[i] + \omega_{i,j}$,  $\mathrm{pre}[j] \gets i$;
        \ENDIF
    \ENDFOR
\ENDFOR
\STATE Build path $\ncalP_t$ from $\mathrm{pre}[{N_{\mathrm{f}} \cdot t +1}]$ to $\mathrm{pre}[0]$;
\ENDWHILE
\STATE {Set lower bound $l\gets t/2$, upper bound $u\gets t$;}
\WHILE{{$l < u $}}
    \IF{$|\ncalP_{\frac{l+u}{2}}| = N_{\mathrm{f}}+1 $}
        \STATE $u \gets (l+u)/2$;
    \ELSE
        \STATE $l \gets (l+u)/2$;
    \ENDIF
\ENDWHILE
\STATE Set $\ncalP \gets \ncalP_{l}$;
\end{algorithmic}
\end{algorithm}

\subsubsection*{Computational complexity of GFA}
The proposed test-and-double binary search requires a number of iterations in the order of $O(\log{n_{\mathrm{t}}})$. 
Each pathfinding process in Algorithm~\ref{alg3}, by traversing the edge set $E$ in a fixed direction for $N_{\mathrm{f}} + 1$ times, finds the maximum weight path in $\ncalG$, ensuring that for all $e_{i,j} \in \ncalP$, no vertex $u$ exists with $\mathrm{sum}[u] + \omega_{u,j} > \mathrm{sum}[j]$. 
For $t$ total time slots, the computational cost of pathfinding in $\ncalG$ is $O(t \cdot {N_\mathrm{f}}^{2})$. Since pathfinding in $\ncalG$ is performed after each test-and-double binary search, the complexity of Algorithm~\ref{alg3} is $O(\log{n_{\mathrm{t}}} \cdot n_{\mathrm{t}} \cdot {N_{\mathrm{f}}}^{2})$. This is significantly more efficient than a traversal search of the original Problem 2, which would have complexity $O(2^{{N_\mathrm{f}}{N_\mathrm{t}}})$.

Compared to the SFA approach, the GFA approach leverages its expanded solution space to enhance NTN content placement efficiency by identifying the maximum-weight path, thereby utilizing fewer time slots and reducing the total content placement time. Quantitative examples of the benefits of this approach are provided in the following section.
\section{Numerical Results}

In this section, we present extensive numerical results to evaluate the performance of the SFA and GFA algorithms, as introduced in Sections~III and IV, respectively.

\subsection{System-level Assumptions}

\subsubsection*{Non-terrestrial network topology}
For the NTN segment, we employ the System Tool Kit (STK) simulator and, unless otherwise stated, we consider a LEO satellite constellation as the one deployed by Starlink. This constellation consists of 1584 satellites distributed in 24 orbits with an inclination of 53$\mathrm{{}^\circ}$. Each satellite creates seven beams with 20\,km beam diameter \cite{SpaceX}. In some cases, we part from this assumption and consider LEO constellations of different densities or a single geostationary Earth orbit (GEO) satellite to study the effect of the NTN topology.

\subsubsection*{Terrestrial network topology}
We select the geographical area with coordinates ([33-39$\mathrm{{}^\circ}$N], [87-93$\mathrm{{}^\circ}$E]), located in western China, as a typical TN scenario where it is challenging to deploy high-density ground network infrastructure. 
We establish six rectangular regions evenly distributed across this area, with each region containing 100 randomly placed TN BSs and a gateway BS connecting all regional BSs using the minimum spanning tree algorithm. 

\subsubsection*{Caching and content popularity}
Each TN BS is configured to cache 50 files, for a total of 30000 files, each with a size of 20\,MB. For the content popularity distribution, we set the skewness parameter to $\alpha_{r}=1$ for all regions $r$ and the regional file correlation coefficient to $\rho=0.8$, unless otherwise stated. In some cases, we vary the values of $\alpha_{r}$ and $\rho$ to study their effect.

\subsubsection*{Non-terrestrial network links}
For NTN links, we set the transmit power to 30\,dBW, the antenna gain at the transmitter to 38.5\,dBi, and the antenna gain at the receiver to 39.7\,dBi.
\blue{We require a minimum elevation angle of 10° to establish an effective NTN link, resulting in a path loss ranging between 173.4 dB and 184.9 dB \cite{3GPP38811,3GPP38821}.}
We assume a 100\,MHz bandwidth for all links, yielding an NTN achievable link rate ranging from 130\,Mbps to 300\,Mbps. 

\subsubsection*{Terrestrial network links}
For TN links, the transmit power is set to 44\,dBm and the antenna gain at both transmitter and receiver to 16\,dBi \cite{3GPP38901}. The delivery radius of each TN BS is set to 2\,km and the minimum inter-BS distance to 0.5\,km \cite{DowFraSha2020}, yielding achievable data rates ranging between 407\,Mbps and 1\,Gbps. 




\begin{table}[!t]
\renewcommand{\arraystretch}{1.3} 
\caption{Main system-level simulation parameters\label{tab:tableparametersNTNTNContentPopularity} }
\centering
\setlength{\tabcolsep}{2mm}{
\begin{tabular}{c|c}

\toprule
\textbf{Parameter} & \textbf{Value} \\ 
\hline 
\multicolumn{2}{c}{\textbf{NTN parameters}}\\ 
\hline 
LEO orbital altitude \cite{SpaceX} & 550\,km \\ \hline 
Number of LEO satellites \cite{SpaceX} & 1584 \\ \hline 
Satellite inclination \cite{SpaceX} & 53$\mathrm{{}^\circ}$ \\ \hline 
Satellite antenna pattern \cite{SpaceX}  & Reflector with circular aperture \\ \hline 
Satellite beam diameter \cite{SpaceX} & 20\,km \\ \hline 
Minimum elevation angle \cite{3GPP38821} & 10$\mathrm{{}^\circ}$ \\ \hline 
Satellite transmit power \cite{3GPP38811} & 30\,dBW \\ \hline 
Transmit antenna gain \cite{3GPP38821} & 38.5\,dBi \\ \hline 
Receive antenna gain \cite{3GPP38821} & 39.7\,dBi \\ \hline 
Link bandwidth & 100\,MHz \\ \hline 

\multicolumn{2}{c}{\textbf{TN parameters}}\\ 
\hline 
Geographical area served & [33-39$\mathrm{{}^\circ}$N], [87-93$\mathrm{{}^\circ}$E] \\ \hline 
Number of geographical regions & 6 \\ \hline 
Size of each region & 1200\,km${}^{2}$ \\ \hline 
Total number of BSs & 600 \\ \hline 
BS coverage radius \cite{DowFraSha2020} & 2\,km \\ \hline 
BS transmit power \cite{3GPP38901} & 44\,dBm \\ \hline 
Transmit and receive antenna gain \cite{3GPP38901} & 16\,dBi \\ \hline 
Link bandwidth & 100\,MHz \\ \hline

\multicolumn{2}{c}{\textbf{Content popularity parameters}}\\ 
\hline 
Content popularity distribution & Zipf distribution \\ \hline 
Number of files cached by TN BS & 50 \\ \hline 
Cache file size & 20 MB \\ \hline 
Cache content skewness & $\alpha_{r}=1$ \\ \hline 
Regional file correlation & $\rho=0.8$ \\ \hline 

\bottomrule
\end{tabular}}
\end{table}

The main system-level parameters of NTN, TN, and content distribution are respectively listed in 
Table~\ref{tab:tableparametersNTNTNContentPopularity}.

\subsection{State-of-the-art Performance Benchmarks}

To assess the performance of our proposed algorithms, we compare the file placement speed of the following five approaches under different system configurations:
\begin{itemize}
\item \emph{The proposed SFA approach}, employing the procedure described in Section~III and Algorithm~1, where we set $\beta=0.5$ to place equal importance on the NTN participation and superiority indicators $\mu^{\textrm{part}}$ and $\mu^{\textrm{sup}}$.
\item \emph{The proposed GFA approach}, which extends the SFA approach, employing the methodology described in Section~IV and outlined in Algorithm~2 and Algorithm~3, respectively for the construction of the NTN placement graph and to determine the global optimal NTN file placement strategy.
\item \emph{A terrestrial network popularity (TNP) approach} \cite{DowFraSha2020}, using NTN links to deliver files, where the latter are ranked in order of conventional TN file popularity. This approach serves as a baseline for NTN utilization, independent of specific file distribution and NTN mobility considerations.
\item \emph{A maximum backhaul traffic (MBT) approach} \cite{HanLiaPen2022}, using NTN links to deliver files, prioritizing those required by at least one BS in each NTN BS coverage footprint. This method focuses on maximizing NTN delivery capacity, without considering the efficiency of file placement on the receiving side.
\item \emph{A standalone terrestrial network (SA-TN) approach}, using only TN links to deploy files, ranked according to conventional TN file popularity. This is considered a baseline scenario when NTN services are not utilized.
\end{itemize}

Table~\ref{tab:tableapproachcomparison} summarizes the main features of the performance benchmarks considered (SA-TN, TNP, and MBT) along with the proposed methods (SFA and GFA). Table~\ref{tab:tableapproachcomparison} also summarizes the performance-complexity tradeoff incurred by each method in the two case studies considered, namely standalone NTN and TN-plus-NTN.
\blue{In the following, we track the edge cache content delivery performance over time, comparatively illustrating the variation in file placement speed across different methods.}

\begin{table*}[!t]
\renewcommand{\arraystretch}{1.5} 
\caption{Summary of the performance benchmarks considered (SA-TN, TNP, and MBT) along with the proposed methods (SFA and GFA) \label{tab:tableapproachcomparison}}
\centering
\setlength{\tabcolsep}{2mm}{
\begin{tabular}{|c|c|c|c|c|c|}

\toprule
\textbf{Approaches} & \textbf{SA-TN}  & \textbf{TNP}\cite{DowFraSha2020}  & \textbf{MBT}\cite{HanLiaPen2022}  &  \textbf{SFA} &  \textbf{GFA} \\ 
\hline 
\textbf{Segment utilized} & Only TN & \makecell{ NTN (Sec.\,V-C, V-E) \\ NTN+TN (Sec.\,V-F)} & \makecell{ NTN (Sec.\,V-E) \\ NTN+TN (Sec.\,V-F)} & \makecell{ NTN (Sec.\,V-C, V-E) \\ NTN+TN (Sec.\,V-F)} & \makecell{NTN (Sec.\,V-D, V-E)\\ NTN+TN (Sec.\,V-F)} \\ 
\hline 
\textbf{Metric maximized} & \textbackslash  & 
\textbackslash & NTN delivery capacity  & File placement speed  & File placement speed \\ 
\hline 
\textbf{Complexity} & \textbackslash & 
\textbackslash & $O({N_{\mathrm{t}}} \cdot {N_{\mathrm{f}}}^{2} )$ & $O({N_{\mathrm{t}}} \cdot {N_{\mathrm{f}}}^{2} )$ & $O(n_{\mathrm{t}}  \log{n_{\mathrm{t}}} \cdot {N_\mathrm{f}}^{2})$  \\ 
\hline 
\textbf{File placement time (slots)} & \textbackslash & \makecell{${N_{\mathrm{t}}}=40$ (NTN)\\ ${N_{\mathrm{t}}}=12$ (NTN+TN)} & \makecell{${N_{\mathrm{t}}}=38$ (NTN)\\ ${N_{\mathrm{t}}}=9$ (NTN+TN)} & \makecell{${N_{\mathrm{t}}}=34$ (NTN)\\ ${N_{\mathrm{t}}}=6$ (NTN+TN)} & \makecell{${n_{\mathrm{t}}}=27$ (NTN)\\ ${n_{\mathrm{t}}}=4$ (NTN+TN)}  \\ 
\hline 
\textbf{Considers content popularity} & \CheckmarkBold & \CheckmarkBold & \CheckmarkBold & \CheckmarkBold & \CheckmarkBold \\ 
\hline 
\textbf{Considers NTN capacity} & \XSolidBold & \XSolidBold & \CheckmarkBold & \CheckmarkBold & \CheckmarkBold \\ 
\hline 
\textbf{Considers TN file distribution} & \XSolidBold & \XSolidBold & \XSolidBold & \CheckmarkBold & \CheckmarkBold  \\ 
\hline 
\textbf{Considers time horizon} & \XSolidBold & \XSolidBold & \XSolidBold & \XSolidBold & \CheckmarkBold \\ 
\bottomrule
\end{tabular}}
\end{table*}

\subsection{Content Placement Speed vs. NTN Topology}

\begin{figure}[t]
\centering
\includegraphics[width=\figwidth]{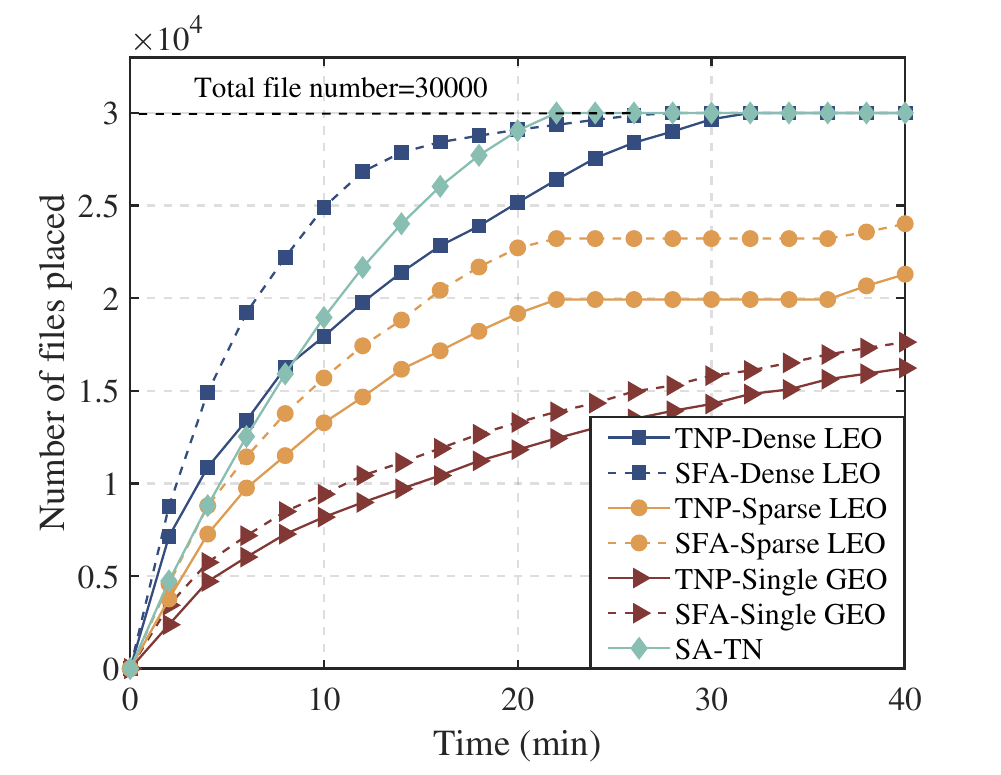}
\caption{Files placed over time by the TNP and SFA approaches for the three NTN segments considered (Dense LEO, Sparse LEO, and Single GEO) and for the SA-TN baseline.
}
\label{fig:numberFilesPlaced_vs_time}
\end{figure}

In Fig. \ref{fig:numberFilesPlaced_vs_time}, we begin by assessing how the topology of the available NTN segment affects the content placement speed. For the time being, we assume all approaches to avail of NTN links only. Specifically, we compare the following scenarios:
\begin{itemize}
    \item A LEO constellation, as outlined in Table~\ref{tab:tableparametersNTNTNContentPopularity}, ensuring uninterrupted coverage of the geographical area under consideration, denoted as \emph{Dense LEO}. 
    \item A LEO constellation with the same parameters as the above but comprising just 11 satellites per orbit, thus only able to cover the geographical area under consideration intermittently, denoted as \emph{Sparse LEO}. 
    \item A setup with a single GEO satellite ensuring uninterrupted coverage but with a lower data rate of 50\,Mbps \cite{DowFraSha2020}, denoted as \emph{Single GEO}.
\end{itemize}
For the above three NTN segments, Fig. \ref{fig:numberFilesPlaced_vs_time} shows the number of files placed vs. time for the proposed SFA algorithm and for the TNP approach. The performance of a SA-TN baseline is also shown for comparison.

Fig. \ref{fig:numberFilesPlaced_vs_time} shows that the proposed SFA algorithm (dashed lines) enhances the file placement speed for all three types of NTN segments considered when compared to a TNP approach (solid lines). The most substantial improvement can be observed in the case of continuous LEO coverage (Dense LEO scenario), during the early stages of content placement. 
In this case, unlike TNP, the SFA approach enables a faster content delivery than the SA-TN.
\blue{These results suggest} that the proposed SFA algorithm is better suited to fully leverage the content delivery capabilities of an NTN segment with respect to an approach that follows conventional TN file popularity. 

Fig. \ref{fig:numberFilesPlaced_vs_time} also demonstrates the advantage provided by a dense LEO constellation (square markers) with respect to a sparser LEO constellation (round markers) and to a single GEO satellite (triangle markers), whose content placement speed is limited by an intermittent coverage and by a limited capacity per geographical area, respectively. Therefore, in the remainder of this section we will focus on comparing the performance of various content placement algorithms when continuous LEO coverage is available, i.e., under a Dense LEO scenario.

\subsection{Content Placement Speed vs. Popularity Distribution}

\subsubsection*{Effect of the popularity skewness}
\begin{figure*}[!t]
\centering
\subfloat[$\rho=0.8, \, \alpha_{r}=0.5$]{\includegraphics[width=56mm]{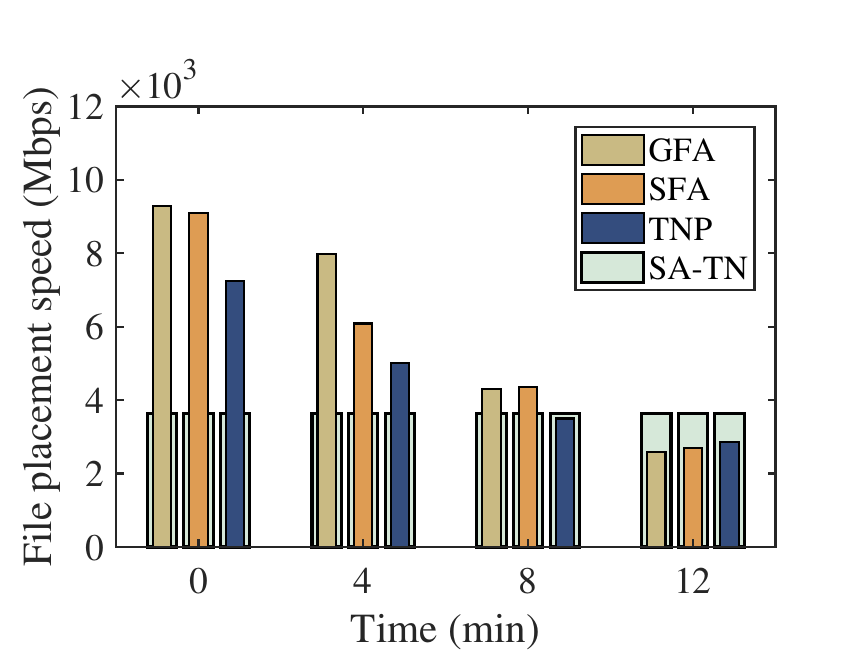}
\label{fig:filePlacementSpeed_vs_alpha_a}}
\hspace*{3mm}
\subfloat[$\rho=0.8, \, \alpha_{r}=1$]{\includegraphics[width=56mm]{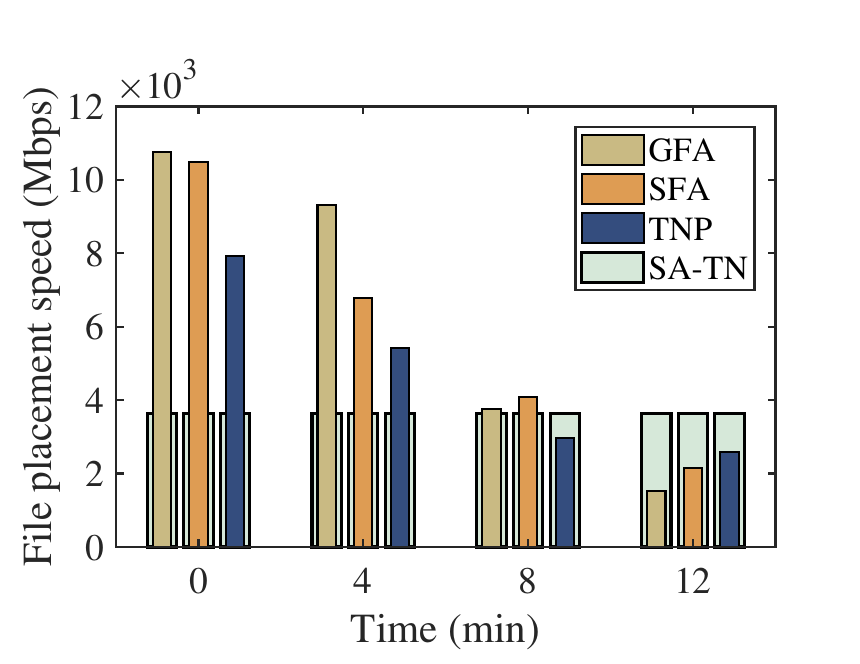}
\label{fig:filePlacementSpeed_vs_alpha_b}}
\hspace*{3mm}
\subfloat[$\rho=0.8, \, \alpha_{r}=1.5$]{\includegraphics[width=56mm]{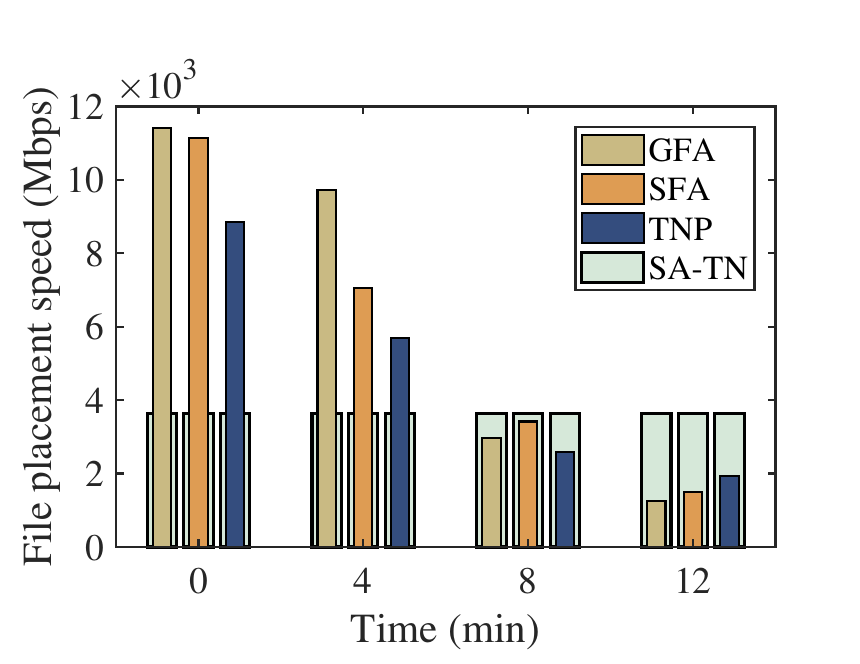}
\label{fig:filePlacementSpeed_vs_alpha_c}}
\captionsetup{justification=justified}
\caption{File placement speed over time for the proposed GFA, the proposed SFA, and the two benchmarks TNP and SA-TN, for different values of the file popularity skewness, $\alpha_{r}$.}
\label{fig:filePlacementSpeed_vs_alpha}
\end{figure*}

In Fig.~\ref{fig:filePlacementSpeed_vs_alpha}, we compare the proposed SFA and GFA algorithms to the TNP and SA-TN approaches in terms of the cumulative data rate provided to all TN BSs versus time. We assume SFA, GFA, and TNP to avail of NTN links only. The data rate provided by SA-TN remains constant over time due to the fixed TN topology. The figure shows such data rates for different values of the file popularity skewness $\alpha_{r}$ ranging from 0.5 to 1.5. 
Fig.~\ref{fig:filePlacementSpeed_vs_alpha} demonstrates that the proposed GFA and SFA schemes exhibit higher file placement speeds than the TNP baseline for all three values of $\alpha_{r}$. In particular, the GFA algorithm outperforms SFA, confirming the effectiveness of a global optimization of the NTN file assignment versus a sequential approach.
As the value of $\alpha_{r}$ increases, so does the proportion of highly popular files and the advantage provided by NTN broadcasting becomes more prominent in the early stages of content delivery. Such advantage vanishes in the later stages, once the most popular content has been delivered, making the TN a better option for content delivery.

\subsubsection*{Effect of the content correlation coefficient}
\begin{figure*}[!t]
\centering
\subfloat[$\alpha_{r}=1, \rho=0.7$]{\includegraphics[width=56mm]{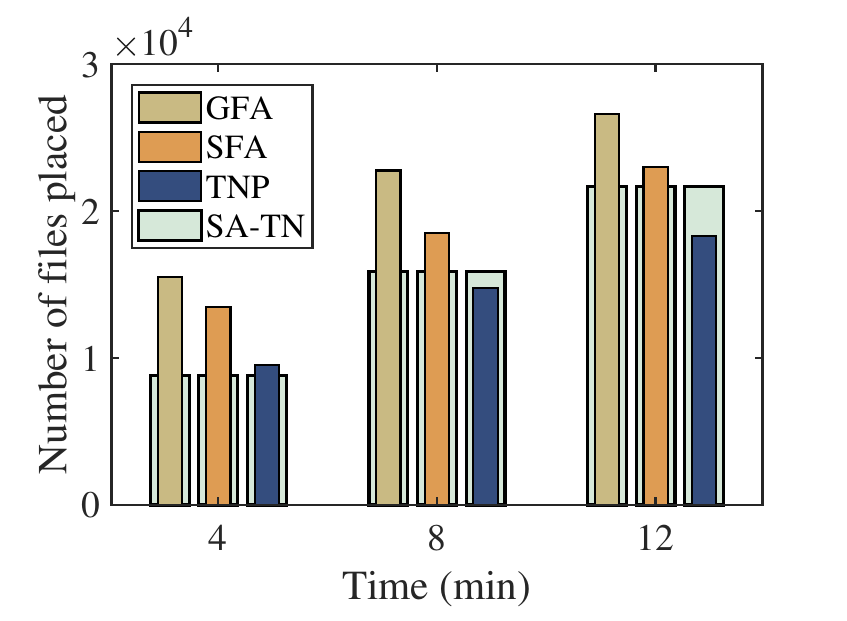}
\label{fig:filePlacementSpeed_vs_rho_a}}
\hspace*{3mm}
\subfloat[$\alpha_{r}=1, \rho=0.6$]{\includegraphics[width=56mm]{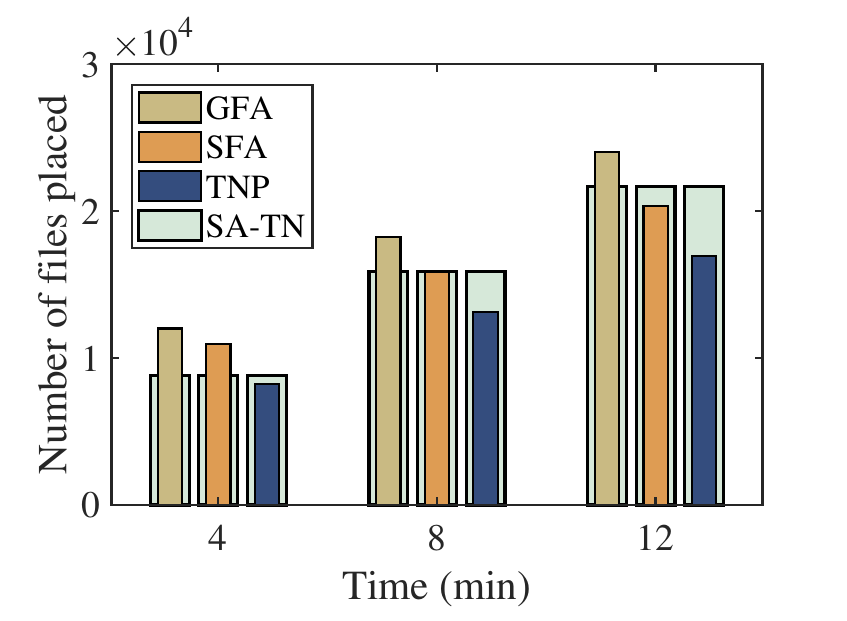}
\label{fig:filePlacementSpeed_vs_rho_b}}
\hspace*{3mm}
\subfloat[$\alpha_{r}=1, \rho=0.5$]{\includegraphics[width=56mm]{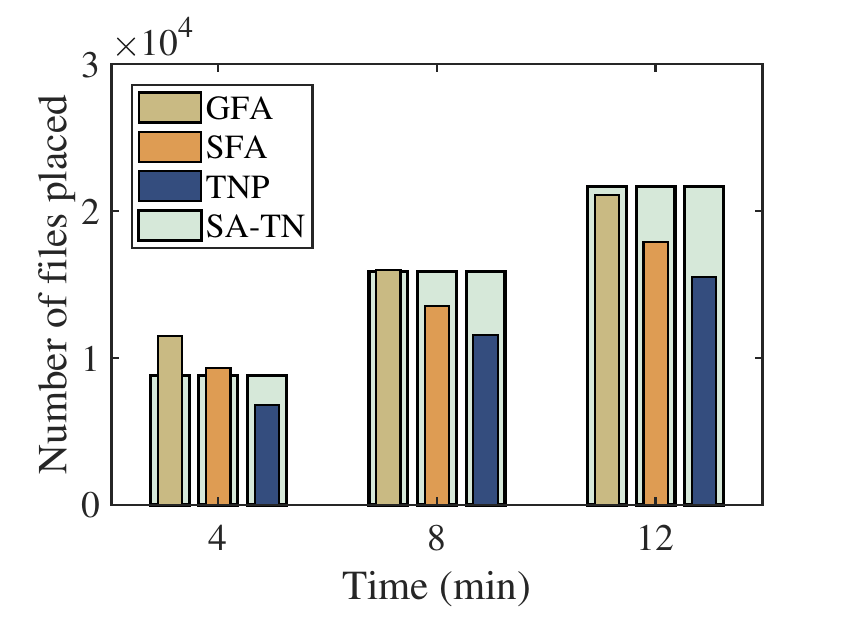}
\label{fig:filePlacementSpeed_vs_rho_c}}
\captionsetup{justification=justified}
\caption{Number of files placed over time for the proposed GFA, the proposed SFA, and the two baselines TNP and SA-TN, for different values of the content popularity correlation across regions, $\rho$.}
\label{fig:filePlacementSpeed_vs_rho}
\end{figure*}

In Fig.~\ref{fig:filePlacementSpeed_vs_rho}, we show the variation in the number of files placed over time for the proposed SFA and GFA algorithms and the TNP approach when employing only NTN links, and we compare it to a SA-TN approach. We also evaluate the impact of the content correlation coefficient $\rho$, with lower values of $\rho$ modeling greater differences in content popularity across regions. 
As expected, as the value of $\rho$ decreases from 0.7 to 0.5, the relative file placement volume of all NTN approaches compared to SA-TN also does. This is due to a greater heterogeneity in file popularity across regions, which reduces the advantage of NTN broadcasting. 
For all values of $\rho$, the proposed GFA and SFA algorithms consistently outperform the TNP baseline, also demonstrating the ability to dynamically adapt to the differences in file popularity among regions.
Once again, the GFA algorithm outperforms the SFA algorithm for different values of $\rho$, since the effectiveness of a global NTN file assignment holds irrespective of the popularity distribution.

\subsection{Comparison with State-of-the-Art Approaches}

\subsubsection*{Effect of NTN coverage fluctuations}

\begin{figure}[t]
\centering
\includegraphics[width=\figwidth]{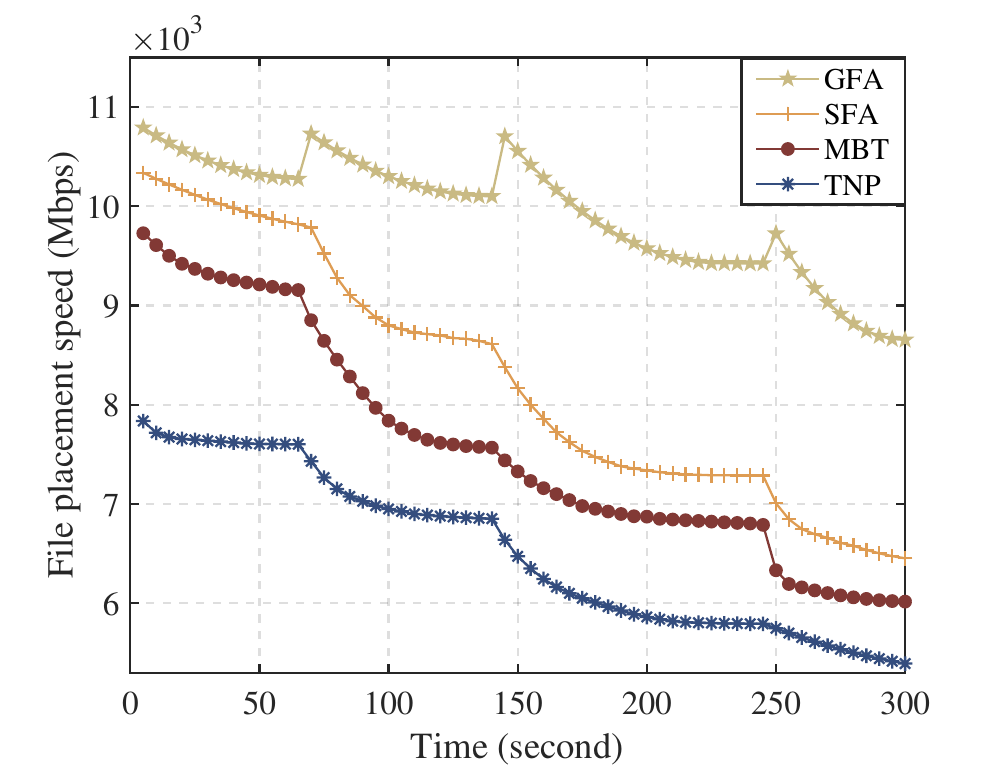}
\caption{File placement speed over the first four time slots for the proposed GFA, the proposed NFA, and the state-of-the-art MBT and TNP approaches.}
\label{fig:filePlacementSpeed_vs_time}
\end{figure}

With the default parameter values of $\rho=0.8$ and $\alpha_{r}=1$, Fig.~\ref{fig:filePlacementSpeed_vs_time} illustrates the variation in cumulative data rate provided to all TN BSs over time. This figure compares the proposed GFA and SFA with the MBT and TNP approaches, all using NTN links exclusively. As per the parameters in Table~\ref{tab:tableparametersNTNTNContentPopularity}, NTN delivery links change with time slots, each lasting from a few seconds to over a minute, covering the first four time slots (300 seconds in total). 
\blue{All four approaches show a decreasing trend in NTN file placement speed over time, both within and across time slots, due to the prioritization of highly popular files initially.} 
The proposed GFA and SFA algorithms consistently outperform MBT and TNP, highlighting their effectiveness in integrating TN file popularity distribution and NTN mobility considerations. 
\blue{Within each curve, the points of discontinuity correspond to time slot switches, indicating changes in NTN coverage status and the initiation of file delivery for the next time slot.}
At these points, only the GFA approach exhibits an increase in file placement speed, arising from its ability to globally deploy each file with the optimal time slot, improving the utilization of NTN delivery capabilities on the entire period. 

\subsubsection*{Gains provided by GFA vs. number of files placed}
\begin{figure}[t]
\centering
\includegraphics[width=\figwidth]{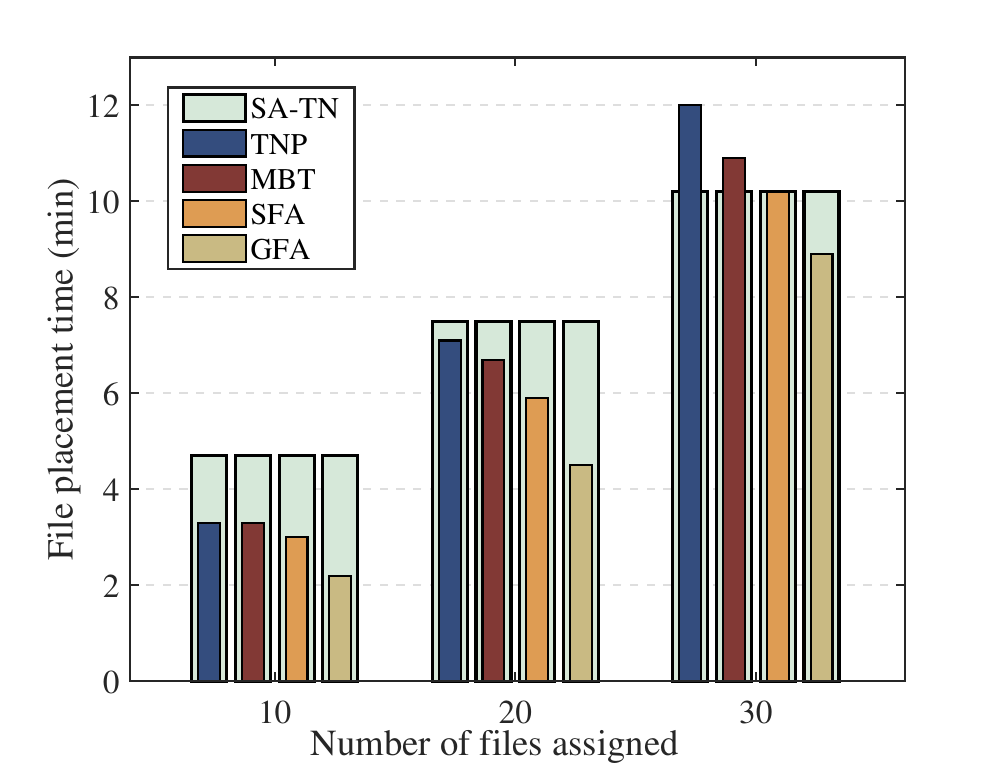}
\caption{File placement time vs. number of files assigned for the proposed GFA, the proposed NFA, the state-of-the-art approaches MBT and TNP, and the baseline SA-TN.}
\label{fig:filePlacementTime_vs_filesAssigned}
\end{figure}

Fig.~\ref{fig:filePlacementTime_vs_filesAssigned} demonstrates the variation in total file placement time for the GFA, NFA, MBT, and TNP approaches, as they arrange 10, 20, and 30 files for NTN placement. The SA-TN baseline is included for context. 
When placing the first 10 files, little difference is observed across the approaches due to the high popularity of these files, which maximizes NTN broadcasting benefits. However, as the number of files increases to 20 and 30, the cumulative time advantage of the GFA becomes increasingly evident. This underscores GFA's ability to provide more efficient and sustained NTN file delivery compared to other strategies. Despite this, the superiority of all NTN-based methods over the SA-TN baseline diminishes with an increasing number of files placed. This trend confirms that TN unicast placement can be more suitable for less popular files.

\subsection{Content Placement Speed in Integrated TN-NTN}

We now evaluate the efficiency of different approaches when both NTN and TN links can be used for content delivery. 

\subsubsection*{Content placement speed of different approaches}

\begin{figure}[t]
\centering
\includegraphics[width=\figwidth]{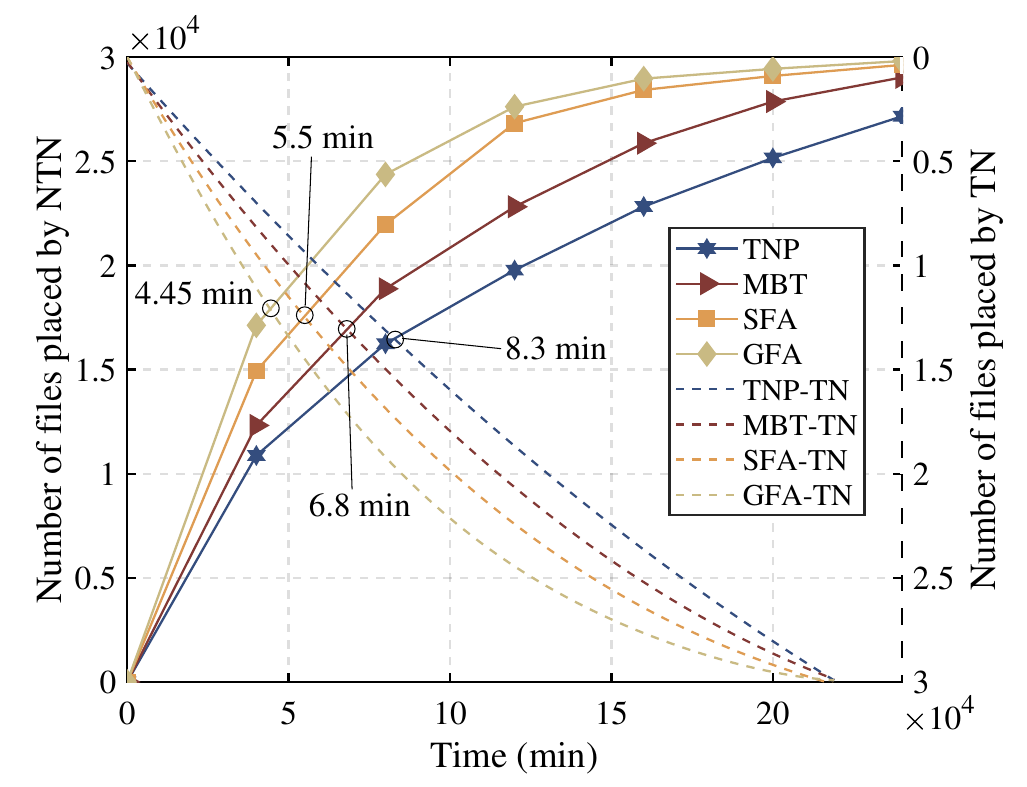}
\caption{Number of files placed vs. time and total placement time via NTN and TN for the proposed GFA and SFA and state-of-the-art MBT and TNP.}
\label{fig:numberFilesPlaced_NTN_TN}
\end{figure}

Fig.~\ref{fig:numberFilesPlaced_NTN_TN} compares the file placement time using the proposed GFA and SFA algorithms against the TNP and MBT approaches. In this scenario, files are deployed through NTN links in descending popularity order (for the four approaches) and through TN links in ascending popularity order. The figure displays the number of files placed over time for each method, using NTN (solid lines) and TN (dashed lines). The intersection of these lines indicates the time to deploy all 30000 files. The SFA algorithm completes file placement in 5.5 minutes, showing a 20\% and 33\% time reduction compared to the MBT and TNP approaches, respectively. The GFA algorithm further reduces this time to 4.45 minutes, achieving 35\% and 47\% faster placement than MBT and TNP.

\subsubsection*{Content placement speed vs. NTN constellation density} 
\begin{figure}[t]
\centering
\includegraphics[width=\figwidth]{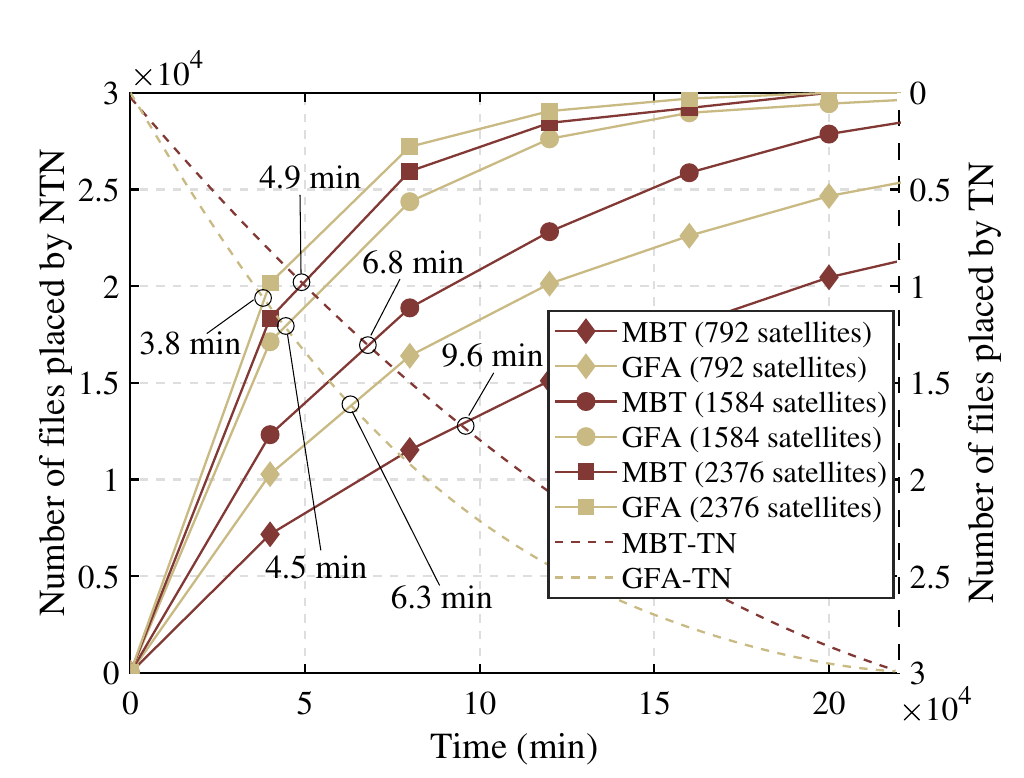}
\caption{ Number of files placed vs. time and total placement time via NTN and TN for the proposed GFA and state-of-the-art MBT with different NTN settings.}
\label{fig:numberFilesPlaced_NTN_TN_vs_numberSatellites}
\end{figure}

We now explore the impact of adjusting the NTN segment's configuration. 
Fig.~\ref{fig:numberFilesPlaced_NTN_TN_vs_numberSatellites} illustrates how varying the number of LEO satellites affect file placement efficiency. We recall that the original LEO constellation consists of 1584 satellites spread across 24 orbits. We now tweak this setup in two ways, namely by adding and subtracting 33 satellites per orbit and resulting in constellations of 792 and 2376 satellites, respectively. We compare the proposed GFA algorithm with the state-of-the-art MBT approach. For all scenarios considered, the proposed GFA algorithm reduces the total content delivery time with respect to the MBT approach by about 20\%. 
Indeed, unlike MBT whose decisions are driven solely by the NTN segment, GFA strategically optimizes the file placement depending on the features of both the TN and NTN segments, demonstrating robustness and capability to adapt to different satellite constellations.

\begin{figure}[t]
\centering
\includegraphics[width=\figwidth]{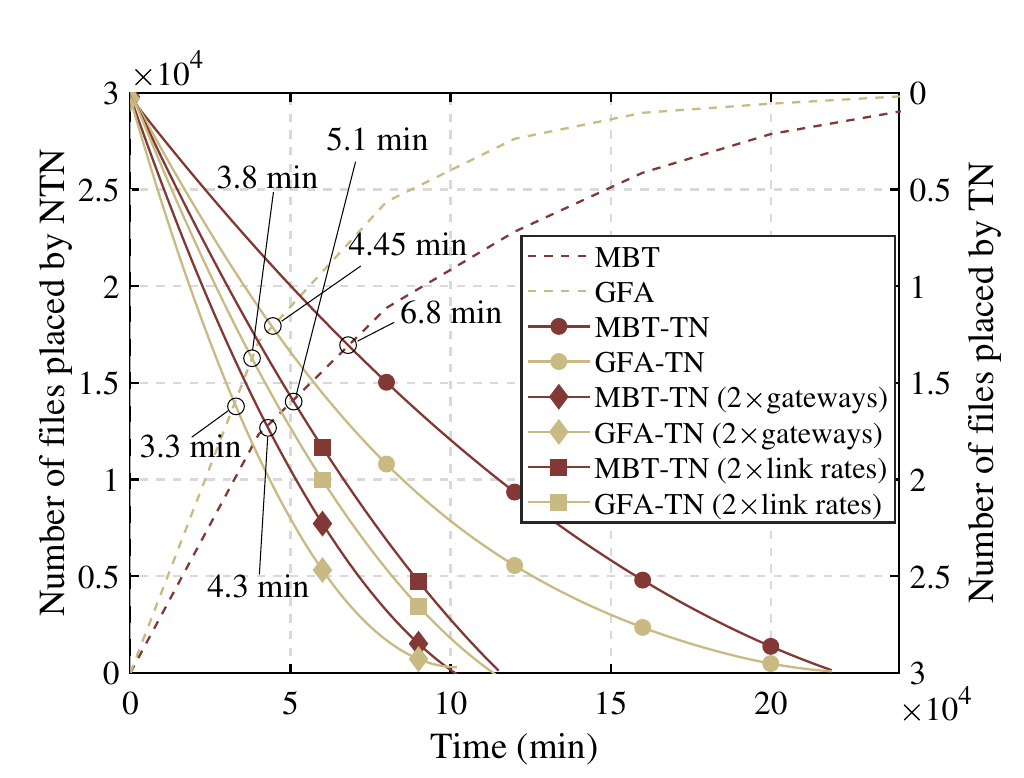}
\caption{Number of files placed vs. time and total placement time via NTN and TN for the proposed GFA and state-of-the-art MBT with different TN settings.}
\label{fig:numberFilesPlaced_NTN_TN_vs_numberRegions_Gbps}
\end{figure}

\subsubsection*{Content placement speed vs. TN topology}

In Fig.~\ref{fig:numberFilesPlaced_NTN_TN_vs_numberRegions_Gbps} we examine the impact of TN link capacity and regional divisions on file placement efficiency. Without altering TN BS cache file requirements or distribution, we show the file placement over time under our original TN settings (circles) and two TN enhancements, namely: (i) increasing the TN link bandwidth from 100\,MHz to 200\,MHz, equivalent to doubling the original rate to between 814\,Mbps and 2\,Gbps (squares), and (ii) increasing the number of evenly divided regions from 6 to 12 (diamonds). The original LEO constellation of 1584 satellites is used for the NTN segment (dashed lines). 
For both GFA and MBT, each of the two TN enhancements nearly halves the TN placement time, for 22 to 10--12 minutes.\footnote{A finer-grained region division (diamonds) marginally outperforms a doubled link rate (squares), because the former not only doubles the number of gateways serving as regional transmission root nodes, but additionally reduces the long-distance hops of each BS from its new and closer gateway.} 
The total TN-plus-NTN delivery time for each case can be observed as the intersection between dashed and solid lines. In all three scenarios, the GFA approach reduces the total content delivery time by about 23\% compared to MBT, demonstrating GFA's efficiency across various TN configurations.

From Fig.~\ref{fig:numberFilesPlaced_NTN_TN} to Fig.~\ref{fig:numberFilesPlaced_NTN_TN_vs_numberRegions_Gbps}, it is evident that integrated TN and NTN can significantly boost file placement speed compared to standalone TN or NTN systems. This increase is due to the effective and opportunistic use of NTN broadcast and TN unicast delivery capabilities. Moreover, different file assignment strategies substantially impact placement speed in integrated networks, with the proposed GFA approach consistently offering over 20\% time savings compared to the MBT approach in diverse scenarios.

\section{Conclusions}

This paper introduced innovative strategies for optimizing wireless edge content placement within integrated terrestrial and non-terrestrial networks. 
Our proposed methods, the sequential file  assignment (SFA) and global file assignment (GFA), focus on the dynamic selection and placement of content via NTN links. These strategies account for the satellite mobility and content popularity distribution, significantly improving content placement speed. The SFA approach begins by assessing the NTN delivery advantage across different content popularity levels, thereby directing sequential content placement through NTN links.
The more advanced GFA method, a graph-based solution, globally allocates content across all delivery slots, thereby maximizing time efficiency in NTN content placement.

Our system-level case studies, using a practical LEO satellite constellation, highlight the advantages of NTN links compared to standalone TN solutions, especially in the initial stages of content placement. These advantages diminish as more popular content is delivered. Additionally, the effectiveness of NTN-based broadcast delivery grows with the correlation of content popularity across different regions. In various scenarios, the GFA approach surpasses the SFA by employing joint optimization across all time slots. Both GFA and SFA methods accelerate placement speed compared to existing methods that do not fully capitalize on dynamic NTN capabilities due to satellite mobility. Remarkably, GFA achieves time savings between 35\% and 47\% in combined NTN and TN placement scenarios.

In this paper, we aimed to advance the understanding of wireless edge content placement optimization through NTN. Our findings emphasized the significant impact of varying NTN delivery capabilities and how optimal content placement should account for content popularity and satellite mobility.




\bibliographystyle{IEEEtran}

\end{document}